\documentclass[10.75pt, a4paper]{article}
\usepackage{titlesec}
\titleformat{\section}
  {\bf\sffamily}
  {\thesection. }
  {5pt}
  {\MakeUppercase}
\renewcommand{\thesection}{\Roman{section}} 

\titleformat{\subsection}
  {\bf\sffamily}
  {\thesubsection. }
  {5pt}{}
  
\renewcommand{\thesubsection}{\Alph{subsection}}

\usepackage[affil-it]{authblk} 
\usepackage{lmodern}

\usepackage[utf8]{inputenc}

\usepackage{geometry}
\usepackage[hidelinks]{hyperref}
\usepackage[citestyle=numeric,style=phys,biblabel=brackets,backend=bibtex,sorting=none,url=false,doi=false, natbib]{biblatex}
\bibliography{bibliography}
\DefineBibliographyStrings{english}{andothers={\itshape et\addabbrvspace al\adddot}}

\usepackage{float}

\usepackage{bm}

\usepackage{abstract}

\usepackage{multirow}

\usepackage[framemethod=tikz]{mdframed}
\usepackage{lipsum}
\usepackage{enumitem}
\usepackage{mdframed}
\usepackage{physics}
\usepackage{setspace}
\usepackage{subfigure}
\usepackage[toc,page]{appendix}

\usepackage[most]{tcolorbox}
\newtcolorbox{myBox}[3][]{
arc=2mm,
lower separated=true,
fonttitle=\bfseries,
colbacktitle=gray!10,
coltitle=black!50!black,
enhanced,
colframe=gray!10,
colback=gray!10,
title=#2,#1}

\usepackage{dirtytalk}
\usepackage{tcolorbox}
\usepackage[font=footnotesize]{caption}
\newtcolorbox{mybox}[1]{colback=green!6!white,colframe=black!75!black,fonttitle=\bfseries,title=#1}
\newtcolorbox{mybox2}{colback=red!5!white,colframe=red!75!black}

\usepackage{pifont}

\usepackage{soul,xcolor}
\setstcolor{red}

\usepackage{xcolor,hyperref}
\hypersetup{
   colorlinks,
   linkcolor={blue!50!black},
   citecolor={blue!50!black},
   urlcolor={blue!80!black}
} 

\definecolor{mycolor}{rgb}{0.122, 0.435, 0.698}

\usepackage{textgreek}

\title{{Predicting Energy Budgets in Droplet Dynamics: A Recurrent Neural Network Approach}}

\author[1]{Diego A. de Aguiar \footnote{diego.aguiar@unesp.br, ORCID: 0000-0001-5159-867X}}
\author[2]{Hugo L. Fran\c{c}a \footnote{franca.hugo1@gmail.com, ORCID: 0000-0002-5361-7704
}}
\author[1]{Cassio M. Oishi\footnote{cassio.oishi@unesp.br, ORCID: 0000-0002-0904-6561}}

\affil[1] {
	Departamento de Matem\'atica e Computa\c c\~ao, Faculdade de Ci\^encias e Tecnologia, Universidade Estadual Paulista ``J\'ulio de Mesquita Filho'', Presidente Prudente, Brazil}
\affil[2]{
    Van der Waals-Zeeman Institute, Institute of Physics, University of Amsterdam, Amsterdam, The Netherlands}
\date{October 2023}

\begin{document}

\begingroup
\sffamily
\date{}
\maketitle
\endgroup

\begin{abstract}
The application of neural network-based modeling presents an efficient approach for exploring complex fluid dynamics, including droplet flow. In this study, we employ Long Short-Term Memory (LSTM) neural networks to predict energy budgets in droplet dynamics under surface tension effects. Two scenarios are explored: droplets of various initial shapes impacting on a solid surface and collision of droplets. Using dimensionless numbers and droplet diameter time series data from numerical simulations, LSTM accurately predicts kinetic, dissipative, and surface energy trends at various Reynolds and Weber numbers. Numerical simulations are conducted through an in-house front-tracking code integrated with a finite-difference framework, enhanced by a particle extraction technique for interface acquisition from experimental images. Moreover, a two-stage sequential neural network is introduced to predict energy metrics and subsequently estimate static parameters such as Reynolds and Weber numbers. Although validated primarily on simulation data, the methodology demonstrates potential for extension to experimental datasets. This approach offers valuable insights for applications such as inkjet printing, combustion engines, and other systems where energy budgets and dissipation rates are important. The study also highlights the importance of machine learning strategies for advancing the analysis of droplet dynamics in combination with numerical and/or experimental data.

\textbf{keywords: Droplets $|$ Numerical solution $|$ Prediction $|$ Surface Tension $|$ LSTM $|$ Energy budget}

\end{abstract}

\section{Introduction}\label{sec:introduction}

Droplet dynamics is a complex phenomenon that covers multiple scientific and engineering domains, requiring a collaborative integration of experimental, analytical, and numerical approaches. Recent advances in these methodologies have increased our understanding of the impact of liquid droplets on rigid surfaces, allowing innovations in fields such as aerospace, energy, materials science, biotechnology, forensics, and the food and cosmetics industries. Notable contributions in this context can be found in \cite{Alexander2006, Howard1994, Nikolopoulos2009, Christophe2016, Shiji2018}. These articles clarify some important aspects of this process such as: the dimensionless numbers and their influence on the maximum spreading of the droplet, different deformation regimes and their consequences in the dynamical energy budget of the system.

Analyzing the energy budgets in droplet dynamics is crucial for understanding fluid behavior and optimizing practical applications. When a droplet approaches a surface or interacts with another droplet, energy transfer occurs between the kinetic, surface, and dissipated components. Studying this interplay not only improves the understanding of fundamental fluid dynamics but also enables the improvement of models based on energy balance, such as those predicting the maximum spreading diameter of a droplet on a rigid surface \cite{Chandra1991, Pasandideh1996, Nicolas1999, Chijioke2005, Vatsal2024}. Recent studies have advanced the understanding of energy-budget dynamics in droplet interactions, as for instance \cite{Jae2016_2,Lennon2023,Qingzhe2021,Zhenyu2017,Karrar2019,Antonov2020,Antoine2021}. These works have practical relevance, particularly in applications as the inkjet printing process \cite{Detlef2022}, where precise control over droplet impact dynamics is essential for achieving desired outcomes. More specifically, the understanding of the collision process in inkjet printing is important, influencing the spatial distribution, size, and morphology of the printed droplets. As droplets traverse through air, dynamical fluid interactions can be observed, such as coalescence, fragmentation, bouncing and other complex behaviors. When droplets impact a substrate, factors such as velocity, impact angle, and material properties of both droplet and substrate influence the outcome. The subsequent spreading of droplets further complicates the process, governing the interaction between ink and substrate and affecting droplet shape \cite{Jens2010}. These challenges highlight the intricate interplay of fluid dynamics in inkjet printing, including droplet collisions, substrate interactions, and spreading dynamics. The current work focuses on two key stages outlined in \cite{Detlef2022}: droplet impact with varying shapes on a substrate and droplet coalescence during collisions.

Over the past decades, the simulation of droplet dynamics has been adopted for advancing our understanding of this complex fluid phenomena \cite{Lorstad2004, Hua2015, Zhuan2022}. Computational frameworks employing the classical finite-element, finite-volume, or finite-difference discretization methods \cite{Xing2007, Robert2010, Stephane2018} are combined with methods to track the deformable moving interface (e.g. front-tracking \cite{Tryggvason2011}) and transient boundary conditions during droplet impact \cite{Yu2005}. Simultaneously, experimental investigations have been adopted to establish different scaling laws in the spreading of droplets \cite{Chandra1991, Nick2014, Fukai1995, Jae2016, Shiji2018, Chijioke2005} and in droplet collisions \cite{Kuo2008, Estrade1999, Qian1997, Qian1997_2, Willis2000}. The synergy between these analytical, experimental, and computational methods enhanced the understanding of droplet interactions \cite{Sander2016, Sikalo2005, Attane2007,Ilia2009}.

More recently, the employment of machine learning-based methods \cite{Song2022, Moussa2022, Jiguo2023, Lap2023, Elham2020, Raihan2023}
for droplet dynamics opens a promising avenue to reduce both time and costs associated with traditional numerical simulations. Notably, most of these works are limited to static data analysis \cite{Shihao2024, Moussa2022, Shengzhi2024, Jiguo2023, Lap2023, Liu2021, Hugo2022, Elham2020, Wenlong2024}, with few exceptions as \cite{Olinka2020, Jae2016_2}. Therefore, the motivation of this study is two-fold. Firstly, we propose a machine learning model based on Long Short-Term Memory (LSTM) \cite{Hochreiter1997} to predict the temporal variation of the energy budget by incorporating dimensionless parameters. Second, through using a method designed for extracting particles from interfaces in digital images we extend our framework to be employed for experimental investigations. With advancements in open-source computational tools (e.g., OpenFOAM, Basilisk) and high-speed imaging techniques, our study highlights the potential of physics-informed machine learning models for predicting and analyzing droplet dynamics during transient regimes. While our current tests focus on numerical results for Newtonian fluids, the non-intrusive concept of this approach, adopting only temporal data, allows its applicability to non-Newtonian fluids. The accurate prediction of energy budgets in complex droplet dynamics demonstrates the robustness and versatility of our methodology, positioning it as a valuable tool for future research.

\section{Methods}\label{sec:methods}

\subsection{Mathematical model and overview of the numerical method} \label{sec:math}

The equations that govern the flow behavior of isothermal and incompressible fluids are the continuity and momentum equations, respectively:
\begin{align}
    & \nabla \cdot \mathbf{u} = 0, \label{navier1}\\
    & { \rho\left( \frac{\partial \mathbf{u}}{\partial t} + \nabla \cdot ( \mathbf{u}\mathbf{u} ) \right)	= - \nabla p  + \nabla \cdot \bm{\tau} },\label{navier0}
\end{align}
where $\mathbf{u}$ and $p$ are the velocity and pressure fields and $\rho$ is the fluid density. The stress tensor $\bm{\tau}$ assumes the form of a Newtonian liquid as
\begin{equation}
    \bm{\tau} = 2 \eta \bm{\dot{\gamma}},
\end{equation}
where $\eta$ is the viscosity of the fluid and $\bm{\dot{\gamma}} = \frac{1}{2} (\nabla \mathbf{u} + (\nabla \mathbf{u})^T)$ is the strain-rate tensor with $\nabla \mathbf{u}$ being the velocity gradient.

A dimensionless version of the previous equations can be obtained by scaling the variables as follows
\begin{equation}
    \label{adimensionalizacao}
    \bar{\mathbf{x}} = \frac{\mathbf{x}}{L}, \hspace{10pt} \bar{t} = \frac{tU}{L}, \hspace{10pt} \mathbf{\bar{u}} = \frac{\mathbf{u}}{U}, \hspace{10pt} \bar{p} = \frac{p}{\rho U^2}, \hspace{10pt} \bar{\boldsymbol{\tau}} = \frac{L \boldsymbol{\tau}}{\eta U},
\end{equation}
using $U$ and $L$ as characteristic velocity and length, respectively. 
Dropping the bars, for convenience, the dimensionless version of equations \eqref{navier1}-\eqref{navier0} is
\begin{align}
    & \nabla \cdot \mathbf{u} = 0, \label{eqAdim2}\\
    & \frac{\partial \mathbf{u}}{\partial t} + \nabla \cdot ( \mathbf{u}\mathbf{u} ) = - \nabla p  + \frac{1}{Re}\nabla^2\mathbf{u}, \label{eqAdim1}
\end{align}
where $Re= \frac{\rho U L}{\eta}$ is the dimensionless Reynolds number.

In our tests, two main boundary conditions are considered: no-slip boundaries, to simulate the impact of a droplet onto a rigid plate, and the free surface condition, for the interface between fluid and the external environment (air). This free surface condition is defined as 
\begin{align}
    \label{eqAdim3}
    \left\{ \begin{array}{ll}
        \mathbf{n} \cdot (\mathbf{T} \cdot \mathbf{n}) = \gamma \kappa, \\
        \mathbf{m} \cdot (\mathbf{T} \cdot \mathbf{n})= 0,
    \end{array} \right.
\end{align}
where $\mathbf{n}$ and $\mathbf{m}$ are, respectively, the normal and tangential vectors to the free surface, $\gamma$ is the surface tension coefficient, $\kappa$ is the interface curvature and $\mathbf{T} = \boldsymbol{\tau} - p\mathbf{I}$ is the total stress tensor. The nondimensionalized forms of (\ref{eqAdim3}) can be written as 
\begin{align}
    \left\{ \begin{array}{ll}
        \mathbf{n} \cdot (\bar{\mathbf{T}} \cdot \mathbf{n}) = \frac{1}{We} \bar{\kappa}, \\
        \mathbf{m} \cdot (\bar{\mathbf{T}} \cdot \mathbf{n}) = 0,
    \end{array} \right.
\end{align}
where $\bar{\kappa} = L \kappa$, $\mathbf{\bar{T}} = 2 \boldsymbol{\bar{\overset{.}{\gamma}}} - Re \cdot \bar{p}\mathbf{I}$ and $We = \frac{\rho U^2 L}{\gamma}$ is the Weber number. In the case of collision between two droplets, an additional dimensionless number is adopted, the impact parameter $B = \frac{2b}{D_1 + D_2}$, where $b$ is the center-to-center relative distance between the droplets and $D_1$, $D_2$ are the diameters for each droplet, as seen in \cite{Karrar2019}.

To solve the governing equations numerically, we have employed the established projection method, utilizing the Helmholtz-Hodge decomposition to decouple velocity and pressure from equations \eqref{eqAdim2}-\eqref{eqAdim1}. In this decomposition, presented in \cite{Chorin1979}, any vector field defined in a region $\Omega$ with smooth boundary $\partial \Omega$ can be decomposed as
\begin{equation}
    \tilde{\mathbf{u}} = \mathbf{u} + \nabla \psi,
    \label{hodge}
\end{equation}
where $\mathbf{u}$ is a vector field such that $\nabla \cdot \mathbf{u} = 0$ and $\psi$ is a scalar field also defined in $\Omega$. 

Initially, the projection method solves the momentum equation (\ref{eqAdim1}) for an intermediate velocity field $\tilde{\mathbf{u}}$. For the purpose of decoupling velocity and pressure, the unknown $p$ is approximated by some known scalar field $\tilde{p}$, resulting in the following equation
\begin{equation}
    { \frac{\partial \tilde{\mathbf{u}}}{\partial t} + \nabla \cdot ( \tilde{\mathbf{u}}\tilde{\mathbf{u}} ) = - \nabla \tilde{p} + \frac{1}{Re}\nabla^2\tilde{\mathbf{u}} },
    \label{navierProj1}
\end{equation}
with appropriate boundary conditions. After solving equation (\ref{navierProj1}) and obtaining the velocity field $\tilde{\mathbf{u}}$, it is still necessary to make sure the continuity equation is satisfied. By combining equations \eqref{hodge} and \eqref{eqAdim2}, the following Poisson equation can be written
\begin{equation}
    \nabla^2\psi = \nabla \cdot \tilde{\textbf{u}}.
    \label{poissonproj}
\end{equation}

Finally, by solving equations (\ref{navierProj1}) and (\ref{poissonproj}) we obtain $\tilde{\mathbf{u}}$ and $\psi$ so that the decomposition (\ref{hodge}) can be used to obtain a final velocity field $\mathbf{u}$ that satisfies both the momentum and continuity equations. Equation (\ref{poissonproj}) is solved with homogeneous Neumann boundary condition on rigid walls, while a third type of boundary condition is used for $\psi$ at the free surface to improve numerical stability. More details about the mathematical formulation and boundary conditions in this method can be found in \cite{Oishi2019}. The discretization process is achieved using the finite-difference method for free surface problems, as detailed in \cite{Oishi2016}. An axisymmetric formulation was adopted for the droplet impact simulations, while droplet collision simulations were performed using a two-dimensional code.

\subsection{Energy budget}

Without gravity, the total energy budget is given by the kinetic energy, surface energy and viscous dissipation. The kinetic energy $E_k$ is defined as
\begin{equation}
    \label{eq:energias1}
    E_k = \int_{V}^{} \frac{1}{2} \rho \left\| \mathbf{u} \right\|^2 dV,
\end{equation}
obtained by integrating over the fluid volume $V$. Numerically, it is calculated as the sum of the kinetic energy inside all cells fully inside a droplet. The surface energy $E_s$ is computed as
\begin{equation}
    \label{eq:energias2}
    E_s = \int_{S}^{} \gamma dS.
\end{equation}
calculated by the integral over the fluid surface $S$. Lastly, based on the viscous dissipation, we can define
\begin{equation}
    \label{eq:energias3}
    E_d = \int_{t}^{}\int_{V}^{} 2 \hspace{0.1cm} \eta \hspace{0.1cm} \left\| \overset{.}{\bm{\gamma}} \right\|^2 dV \hspace{0.1cm} dt,
\end{equation}
where $\int_{V}^{} 2 \hspace{0.1cm} \eta \hspace{0.1cm} \left\| \overset{.}{\bm{\gamma}} \right\|^2 dV$ is called the viscous dissipation rate. 

Conservation of energy states that the total sum of all the energies ($E_t$) in the system must remain constant through time, that is
\begin{equation}
    \label{eq:energias4}
    E_t(t) = E_k(t) + E_s(t) + E_d(t) = E_t(0).
\end{equation}
In order to keep our investigation nondimensional, we also rescale the energies \eqref{eq:energias1}-\eqref{eq:energias3} and work with the following quantities
\begin{align}
    \bar{E}_k & = Re We \int_{\bar{V}}^{} \frac{1}{2} \left\| \bar{u} \right\|^2 d \bar{V}, \label{eq:energies_nondim1}\\
    \bar{E}_s & = Re \int_{\bar{S}}^{} 1 \hspace{0.1cm} d \bar{S}, \label{eq:energies_nondim2} \\
    \bar{E}_d & = We \int_{\bar{t}}^{}\int_{\bar{V}}^{} 2 \left\| \bar{\overset{.}{\bm{\gamma}}} \right\|^2 d \bar{V} \hspace{0.1cm} d\bar{t}. \label{eq:energies_nondim3}
\end{align}
The focus of this work is to use geometric time series data extracted from numerical simulations to generate time series predictions for each energy component. From now on, we omit the bars over all nondimensional variables for convenience.

%






\subsection{Problem descriptions and data preparation}

\begin{figure}[H]
    \centering
    \includegraphics[width=1.0\textwidth]{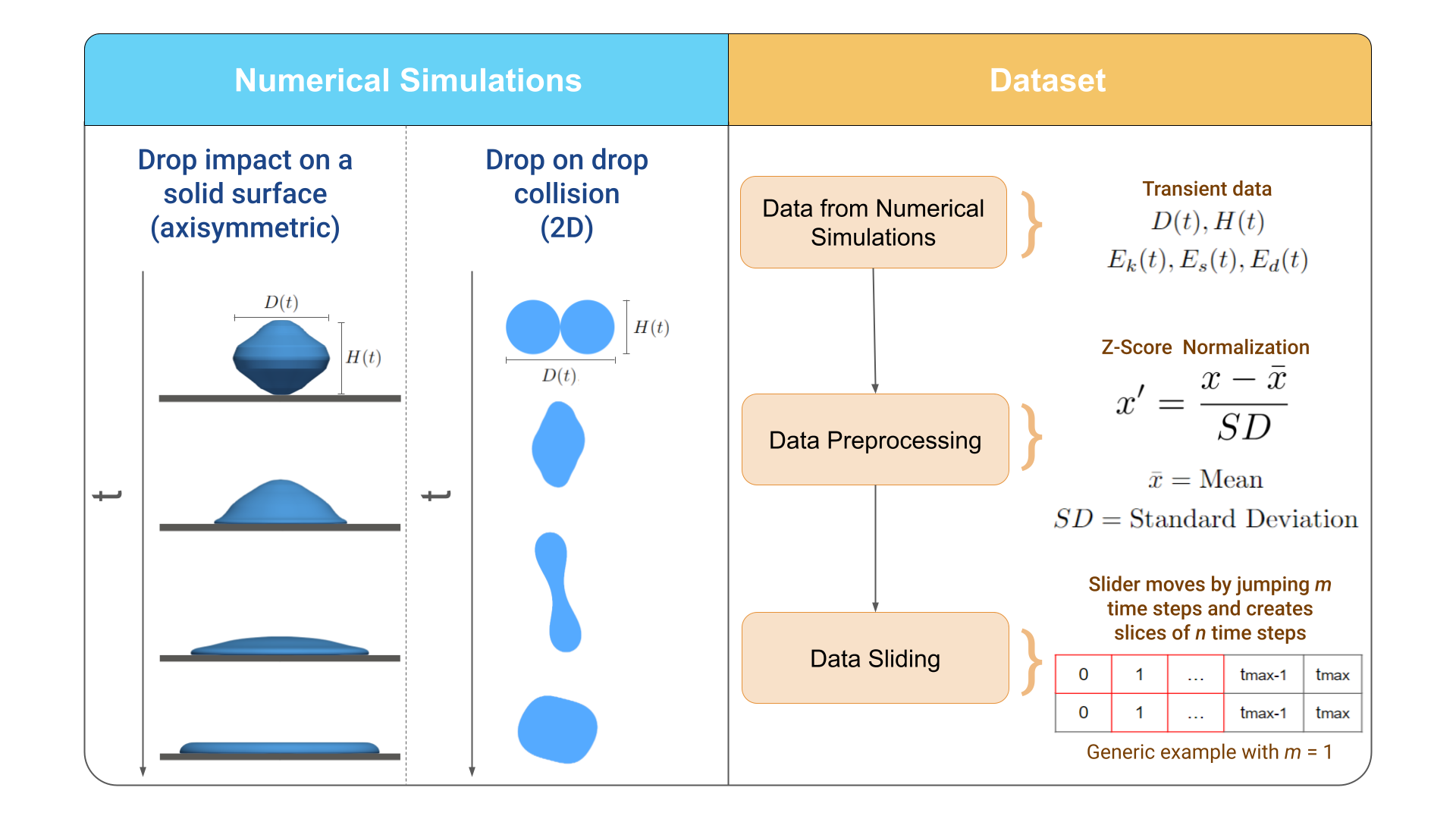}
    \caption{Dataset is extracted from numerical simulations of two droplet dynamic problems.}
    \label{fig:organizacao_1}
\end{figure}

To apply a neural network, it is essential to create a sufficiently representative training dataset. For this purpose, numerical simulations are employed to generate two different flow regimes: (i) droplet impact on a solid surface, to generate diverse spreading diameters with different initial droplet shapes, and (ii) binary droplet collision, to test the capacity of the model to predict features that oscillate through time. Both problem descriptions are represented in Fig. \ref{fig:organizacao_1} (left) while the schematic to obtain the dataset is also described in the same figure (right). Further details about the codes used to generate the datasets can be found in \cite{Oishi2019, Hugo2022_2}.

In order to generate the datasets, each case is executed with varying flow parameters, and/or initial droplet shapes. In the droplet impact scenario, we vary the droplet shape by extracting interfaces from experimental images \cite{Qingzhe2021}. For binary droplet collisions, we induce initial geometrical variations through the impact parameter B. In both cases, we also vary the Reynolds and Weber numbers. In summary, samples within the dataset can be defined as individual data points, each characterized by a specific combination of features encompassing dimensionless parameters ($Re$ and $We$) with initial geometrical droplet characteristics, as will be detailed in the following sections.

\subsubsection{Droplet spreading dataset}

\begin{figure}[H]
    \centering
    \includegraphics[scale=0.5]{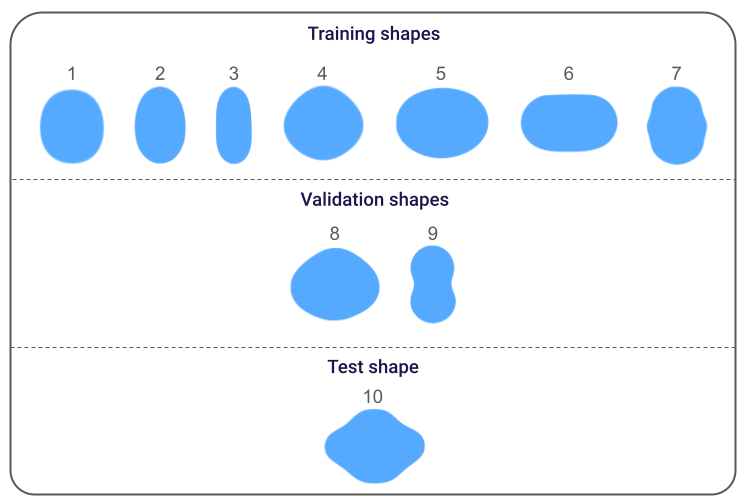}
    \caption{Different initial droplet shapes and the train-validation-test splits. The more complex shapes were specifically chosen for the validation and test sets.}
    \label{fig:conjunto1_1}
\end{figure}

Our first dataset is inspired by droplet shapes from experiments presented by \cite{Qingzhe2021}. To test the efficiency of the machine learning model, the shapes were selected to ensure that the training, validation and test subsets would have sufficiently different shapes, as shown in Fig. \ref{fig:conjunto1_1}. 

An interface extraction method was implemented in order to obtain these shapes from experimental images, as described in Appendix \ref{ap:extraction}. The initial diameter and height of each experimental shape, as presented in Table \ref{tab:tamanhos}, are selected in such a way that each droplet has approximately the same volume. 

\begin{table}[H]
    \begin{center}
        \begin{tabular}{ |p{2cm}||p{3cm}|p{3cm}| }
            \hline
                Shape & $D(0)$ (in mm) & $H(0)$ (in mm) \\
            \hline
                1 & 3.92 & 3.98 \\
                2 & 3.6771 & 4.64 \\
                3 & 2.62 & 4.64 \\
                4 & 4.64 & 3.6829 \\
                5 & 4.6086 & 3.1343 \\
                6 & 4.8143 & 2.5714 \\
                7 & 3.8229 & 4.24 \\
                8 & 4.5057 & 3.3857 \\
                9 & 3.5571 & 5.0086 \\
                10 & 4.6343 & 3.6229 \\
            \hline
        \end{tabular}
        \caption{The initial spreading diameter and height for each droplet shape.}
        \label{tab:tamanhos}
    \end{center}
\end{table}

The parameters chosen to configure the numerical simulations are presented in Table \ref{tab:conjunto1}. For each set of parameters and one droplet shape, a numerical simulation is run to generate one sample. The Reynolds and Weber numbers were arbitrarily defined by varying the impact velocity, with the characteristic length $L$ defined as the initial droplet diameter $D(0)$. In total, 132 simulations were performed, each run until a maximum time of $t_{\text{max}} = 50$, with 1000 equally spaced snapshots extracted per simulation.

\begin{table}[H]
    \begin{center}
        \begin{tabular}{ |p{3cm}||p{3cm}|p{3cm}| }
            \hline
            \multicolumn{3}{|c|}{Spreading dataset} \\
            \hline
                Number of shapes & Re & We \\
            \hline
                10 & 1.5063 & 240.625 \\
                10 & 4.5188 & 2165.625 \\
                10 & 5.875 & 212.5 \\
                10 & 10.2438 & 184.375 \\
                10 & 17.6250 & 1912.5 \\
                10 & 20.4375 & 118.75 \\
                10 & 24.8063 & 90.625 \\
                9 & 30.6313 & 53.125 \\
                10 & 30.7313 & 1659.375 \\
                7 & 35 & 25 \\
                10 & 61.3125 & 1068.75 \\
                9 & 74.4188 & 815.625 \\
                9 & 91.8938 & 478.125 \\
                8 & 105	& 225 \\
            \hline
        \end{tabular}
        \caption{Number of samples for each group of parameters used in the spreading simulations.}
        \label{tab:conjunto1}
    \end{center}
\end{table}

\begin{figure}
\centering
    \subfigure[]{
        \includegraphics[width=0.47\textwidth]{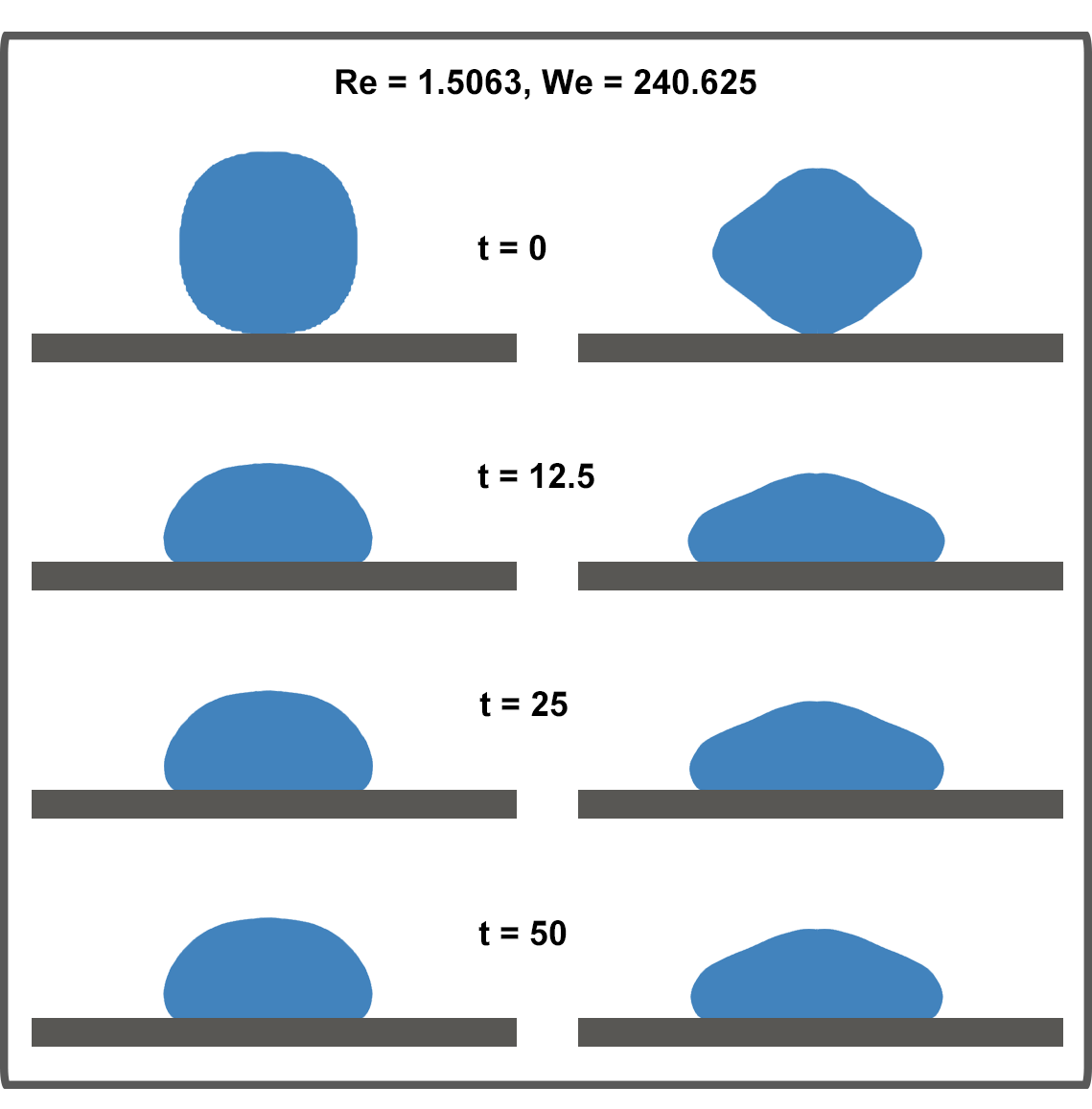}
        \label{fig:conjunto1_2a}
    }
    \subfigure[]{
        \includegraphics[width=0.47\textwidth]{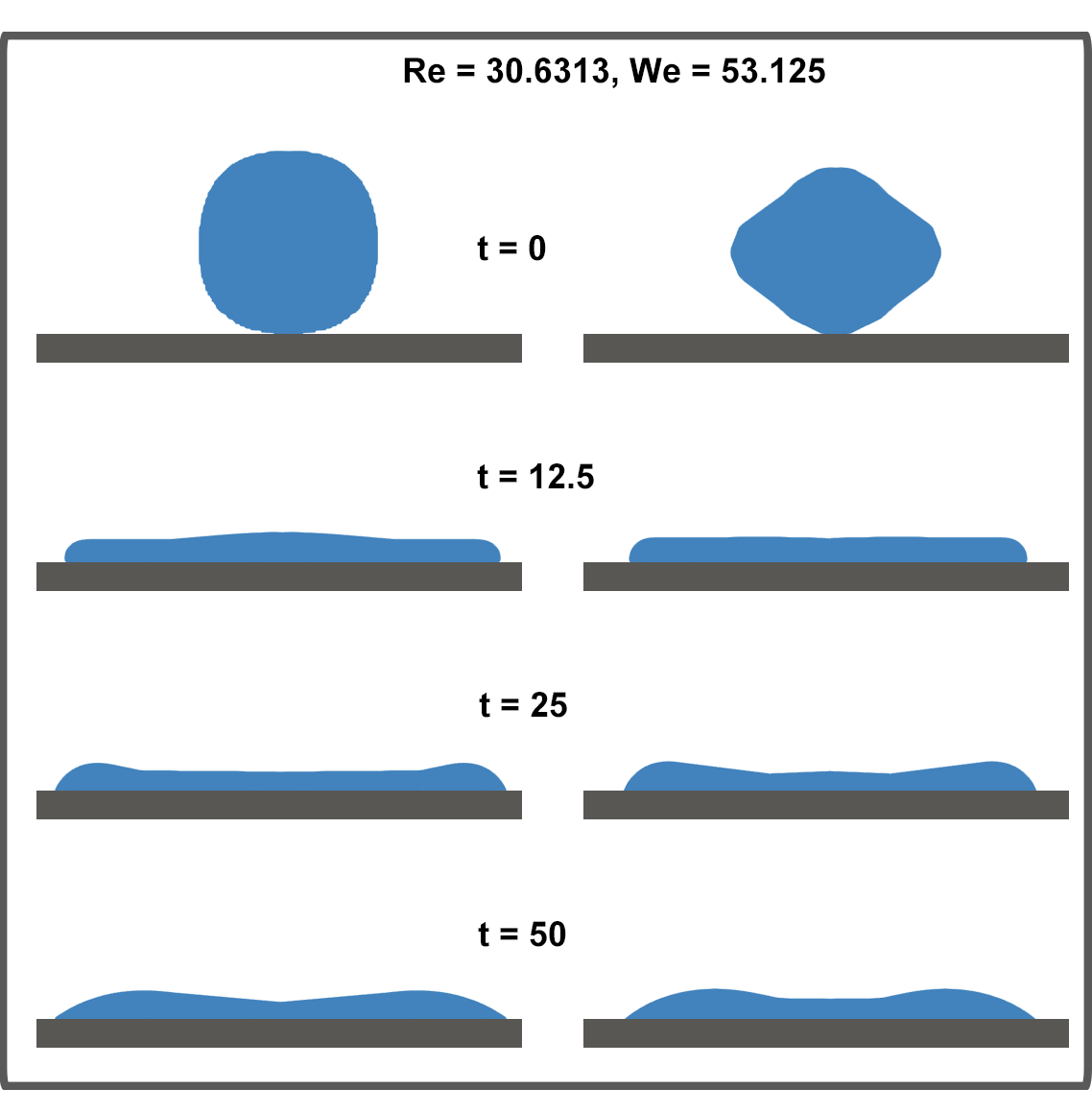}
        \label{fig:conjunto1_2b}
    }
    \subfigure[]{
        \includegraphics[width=0.47\textwidth]{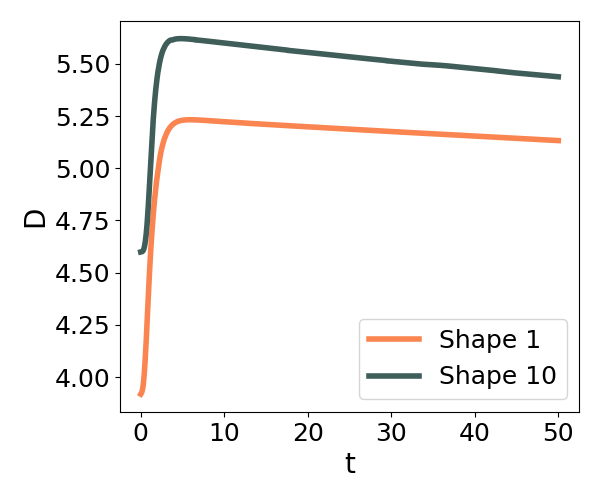}
        \label{fig:conjunto1_2c}
    }
    \subfigure[]{
        \includegraphics[width=0.47\textwidth]{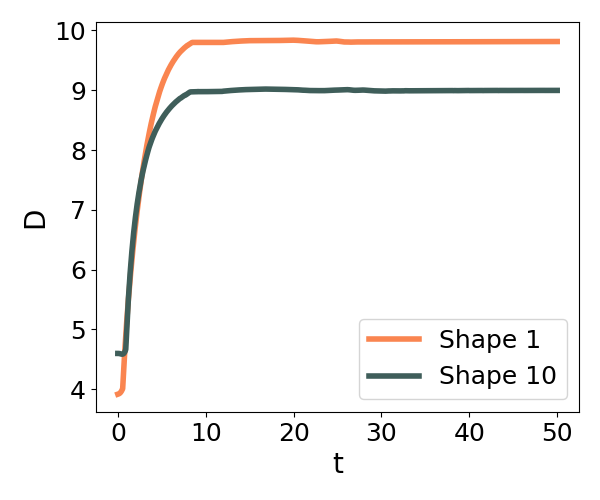}
        \label{fig:conjunto1_2d}
    }
    \caption{Effect of drop shape (shapes 1 and 10 from Fig \ref{fig:conjunto1_1}) on impact when varying Re and We: morphology for the first (a) and second (b) set of parameters and diameter time variation for both shapes from the first (c) and second set (d).}
    \label{fig:conjunto1_2}
\end{figure}

To visualize the spreading behavior while varying Re and We, Fig. \ref{fig:conjunto1_2} illustrates the droplet morphology and the temporal evolution of the diameter for shapes 1 and 10. As depicted in this figure, increasing the Reynolds number results in a greater spread for both shapes. This behavior is quantified by the subplots in Fig. \ref{fig:conjunto1_2} depicting the time variation of the droplet diameter. Figs. \ref{fig:conjunto1_2a} and \ref{fig:conjunto1_2b} illustrate snapshots comparing two shapes for two different sets of dimensionless numbers, while Figs. \ref{fig:conjunto1_2c} and \ref{fig:conjunto1_2d} compare the evolution of the spreading diameter through time for both shapes. Figs. \ref{fig:conjunto1_2a} and \ref{fig:conjunto1_2c} demonstrate that at lower Reynolds numbers, the droplets spread more slowly compared to the second test involving higher $Re$.

\subsubsection{Collision dataset}

The second dataset consists of two drops of 1mm in diameter colliding horizontally. After the collision, the droplets merge and generate both horizontal and vertical oscillations over time, creating time series that are more challenging to predict and that can be used to test the efficiency of the machine learning model.

To define the numerical simulation parameters for each sample, cases similar to those presented in the experimental coalescence map from \cite{Karrar2019} were selected by varying the relative impact velocity, with the characteristic length $L$ defined as the initial diameter of one of the droplets. For each set of Reynolds and Weber numbers, 20 simulations are run by varying the impact parameter B in the intervals shown in Table \ref{tab:conjunto2}. In total, 72 simulations were performed, following the same setup as before, with each running until $t_{\text{max}} = 50$ and producing 1000 equally spaced snapshots.

\begin{table}[H]
    \begin{center}
        \begin{tabular}{ |p{2cm}||p{3cm}|p{3cm}|p{3cm}| }
            \hline
            \multicolumn{4}{|c|}{Collision dataset} \\
            \hline
                $Samples$ & Re & We & B (interval) \\
            \hline
               20 & 74.451 & 22.0 & $[0.0, 0.4]$ \\
               20 & 85.4788 & 29.0 & $[0.0, 0.4]$ \\
               9 & 97.8478 & 38.0 & $[0.0, 0.2316]$ \\
               5 & 104.0863 & 43.0 & $[0.0211, 0.1684]$ \\
               7 & 109.9715 & 48.0 & $[0.0421, 0.1684]$ \\
               6 & 115.5573 & 53.0 & $[0.0421, 0.2105]$ \\
               5 & 120.8853 & 58.0 & $[0.0, 0.2316]$ \\
            \hline
        \end{tabular}
        \caption{Number of samples for each group of parameters used in the collision simulations.}
        \label{tab:conjunto2}
    \end{center}
\end{table}

\begin{figure}
\centering
    \subfigure[]{
        \includegraphics[width=0.3\textwidth]{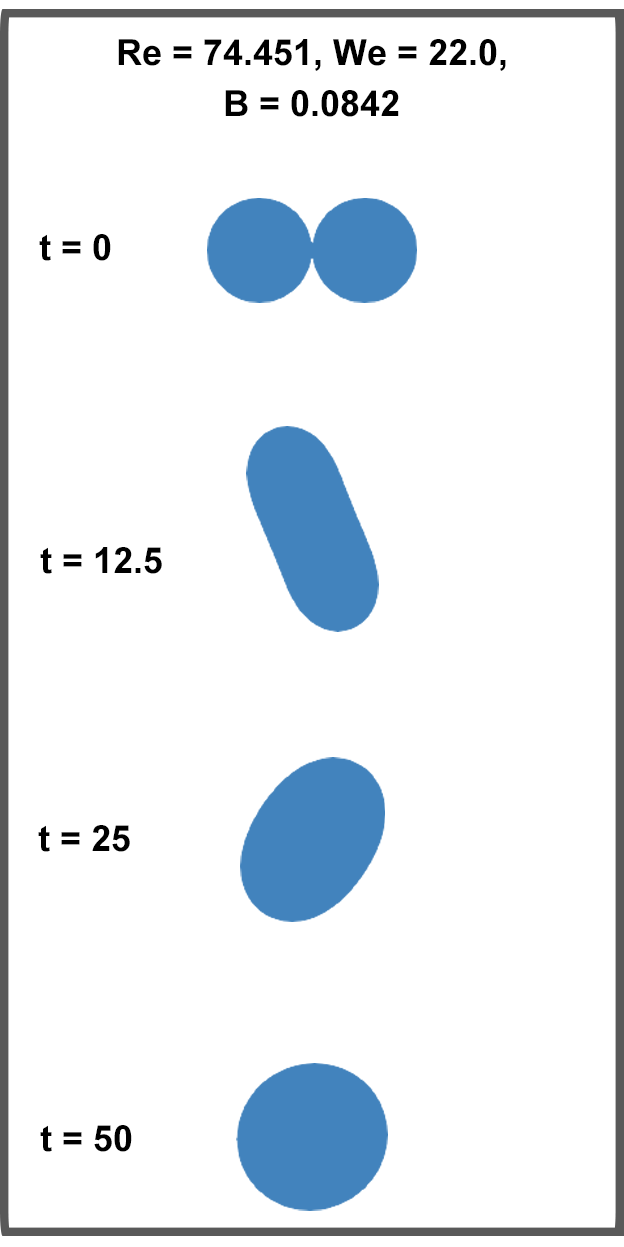}
        \label{fig:conjunto2_1a}
    }
    \subfigure[]{
        \includegraphics[width=0.3\textwidth]{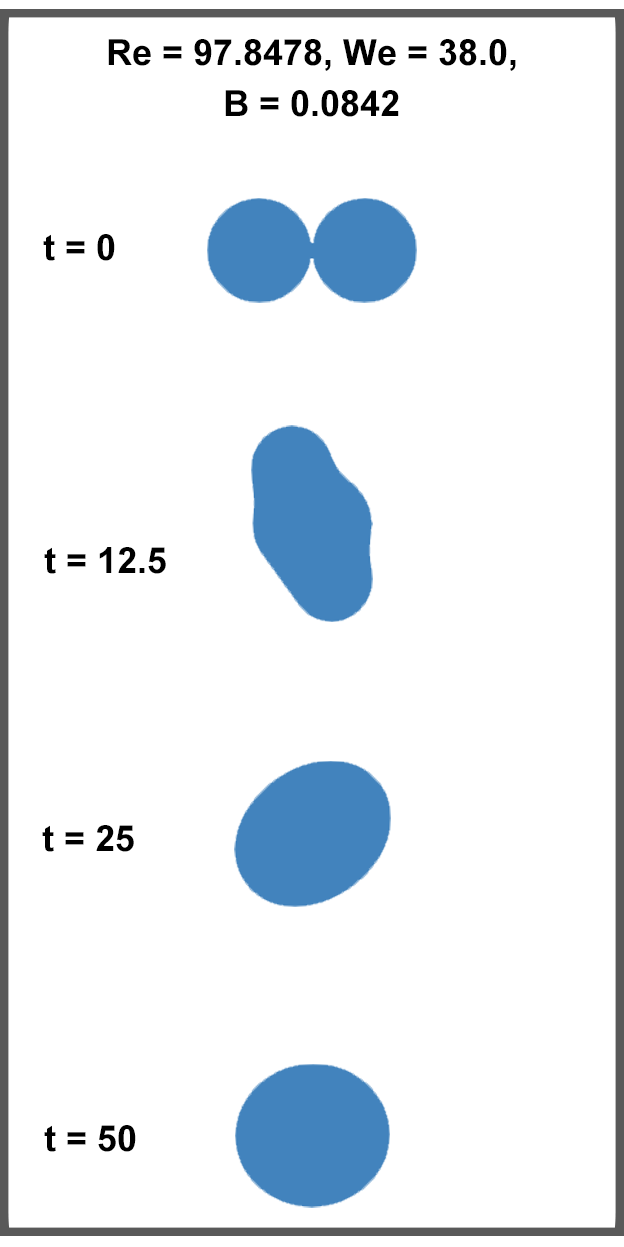}
        \label{fig:conjunto2_1b}
    }
    \subfigure[]{
        \includegraphics[width=0.3\textwidth]{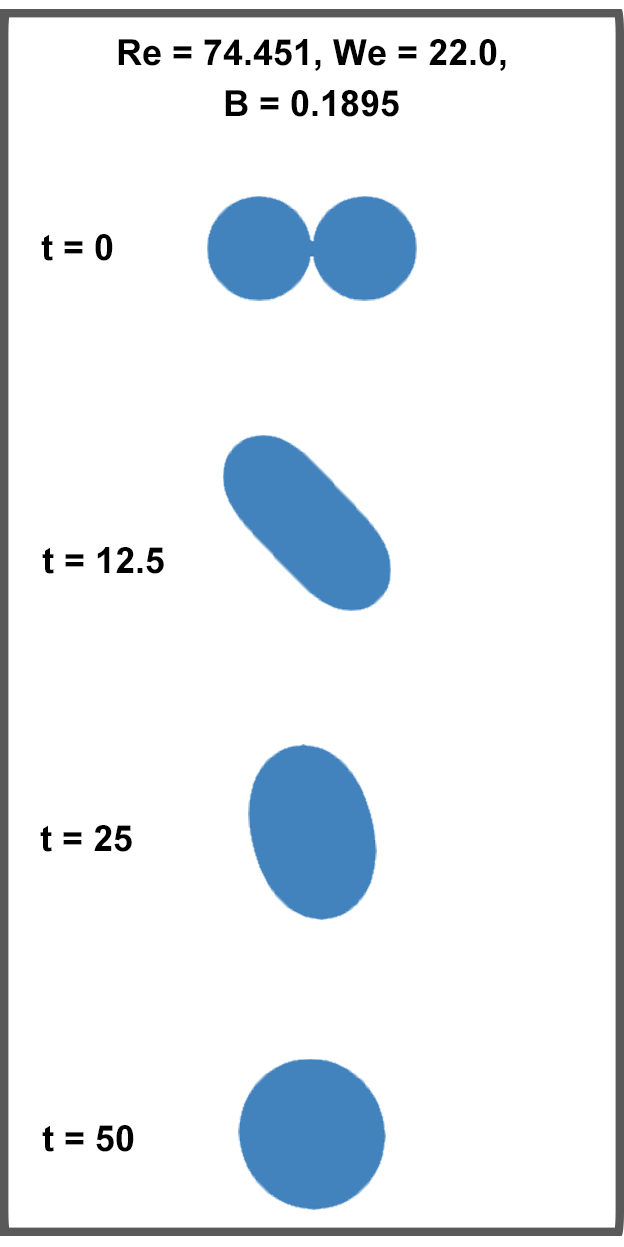}
        \label{fig:conjunto2_1c}
    }
    \subfigure[]{
        \includegraphics[width=0.3\textwidth]{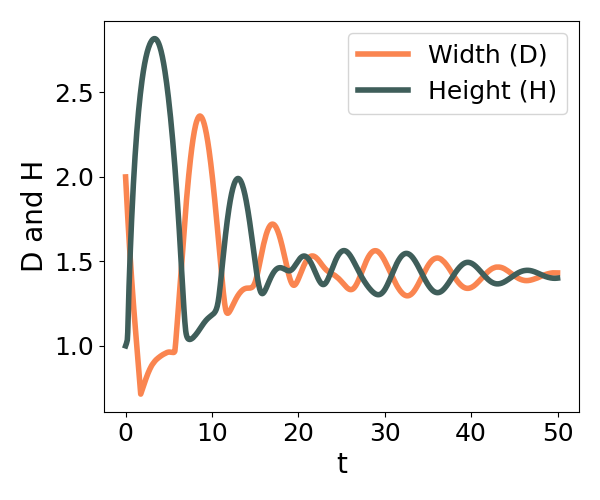}
        \label{fig:conjunto2_1d}
    }
    \subfigure[]{
        \includegraphics[width=0.3\textwidth]{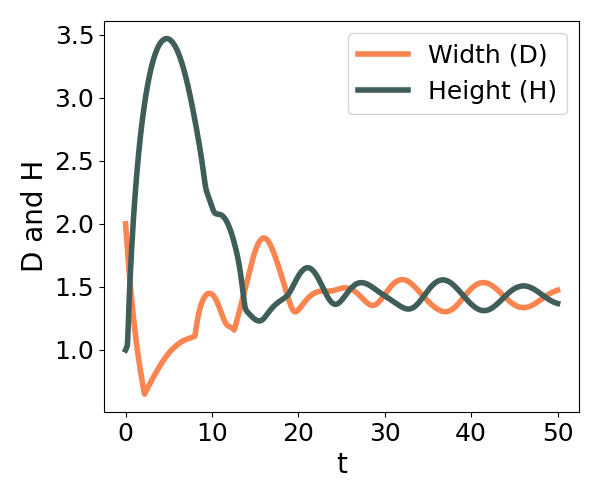}
        \label{fig:conjunto2_1e}
    }
    \subfigure[]{
        \includegraphics[width=0.3\textwidth]{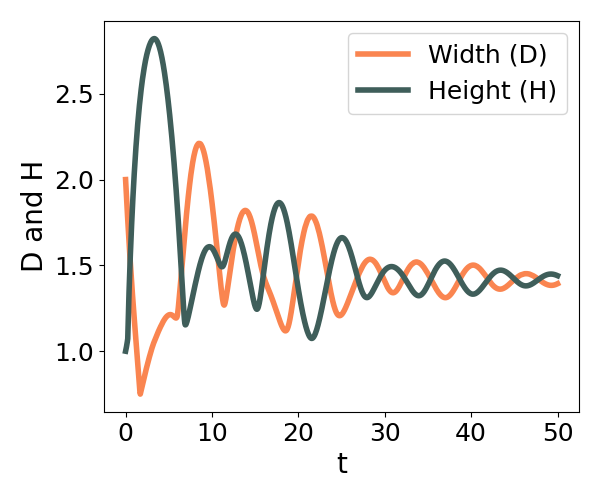}
        \label{fig:conjunto2_1f}
    }
    \caption{Droplet collision dynamics with different impact parameters and varying Re and We: morphology for the first (a), second (b) and third (c) set of parameters and time variation in diameter $D$ and height $H$ for the first (d), second (e) and third set (f).}
    \label{fig:conjunto2_1}
\end{figure}

The effect of parameters on horizontal and vertical elongations after impact is depicted through simulation snapshots in Fig. \ref{fig:conjunto2_1}. Figs. \ref{fig:conjunto2_1a}, \ref{fig:conjunto2_1b}, and \ref{fig:conjunto2_1c} illustrate snapshots of two coalescing droplets on three different sets of dimensionless numbers, with pairs \ref{fig:conjunto2_1a} and \ref{fig:conjunto2_1b} varying only the Reynolds and Weber numbers, and \ref{fig:conjunto2_1a} and \ref{fig:conjunto2_1c} varying only the impact parameter. For Figs. \ref{fig:conjunto2_1d}, \ref{fig:conjunto2_1e}, and \ref{fig:conjunto2_1f}, we compare the time series of diameter and height oscillation for each example.

\subsection{Machine learning model}

In this work, we propose a machine learning model based on a recurrent neural network to predict the temporal variation of the energy budget. Given the complex nature and the underlying temporal dependency of data in this work, we have adopted the Long Short-Term Memory (LSTM) model\cite{Hochreiter1997}. With a geometric input data set, the model learns a mapping to produce temporal predictions in terms of the energy budget. The term ``geometric'' refers to the time variation of the droplet diameter during spreading on a surface or the diameter and height of the coalesced droplet after the collision of two droplets. Incorporating dimensionless parameters such as the Reynolds and Weber numbers, as well as geometrical quantities as the collision impact parameter, enhances the accuracy of predictions for the transient behavior of energy components.

Since the transient features in the extracted data have a certain degree of time-correlation, it would be beneficial to employ a neural network architecture that can capture it effectively. For this purpose, a Recurrent Neural Network \cite{Jain1999}, which is effective at finding correlations on data sequences, was considered as an option.

The drawback of a simple RNN architecture is the likely introduction of the vanishing gradient problem \cite{Seol2021}, considering that our data has a large number of time steps. To address this issue, we selected the LSTM architecture \cite{Ralf2019}, a variant of RNN designed to efficiently learn long-term dependencies in data.

\begin{figure}[H]
    \centering
    \includegraphics[width=0.95\textwidth]{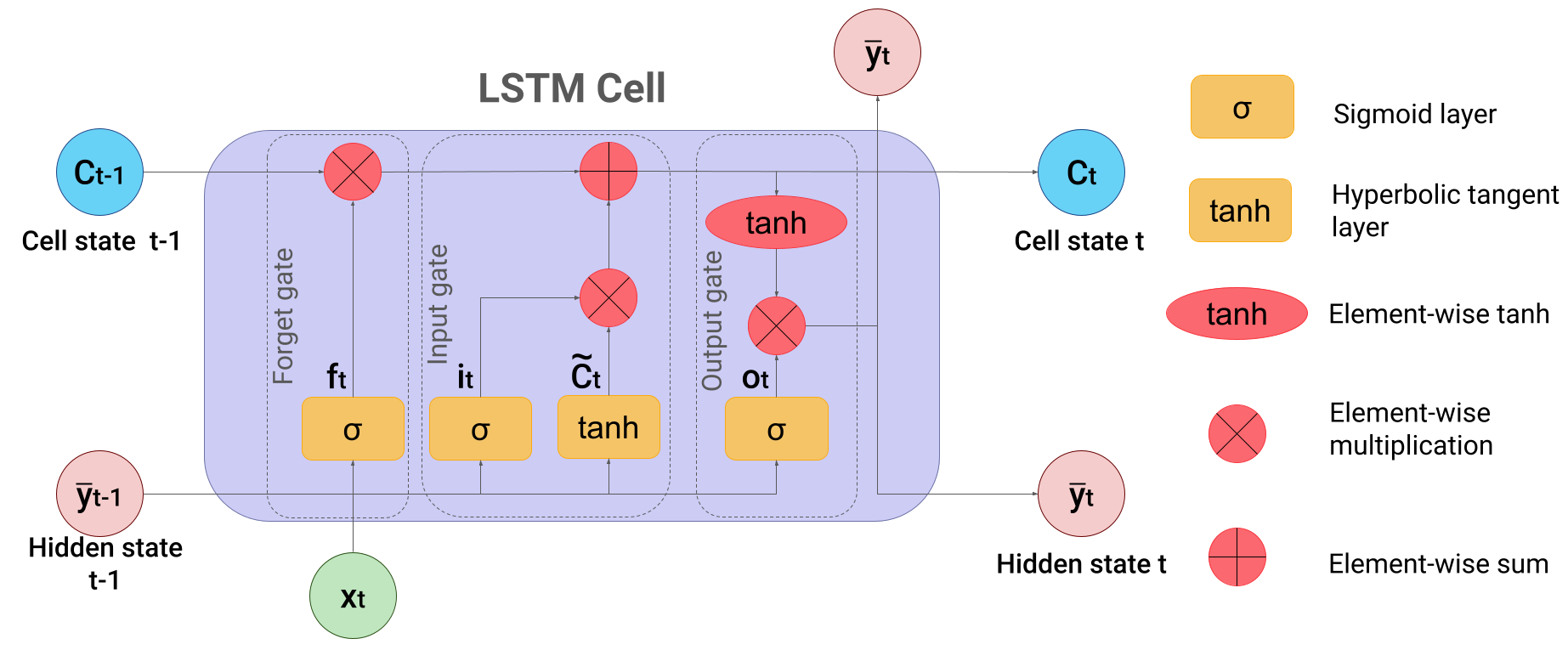}
    \caption{LSTM Cell structure.}
    \label{fig:estrutura_lstm}
\end{figure}

The LSTM architecture introduces two new concepts to increase its capacity to deal with more time steps: a cell state to capture long-term information and three gates that control the intensity of the data passing through them. The structure of an LSTM cell and all of its additional concepts are represented in Fig. \ref{fig:estrutura_lstm}.

Unlike the hidden state, which mostly carries information from the previous time step, the cell state is capable of passing data through multiples cells with minimal reductions and/or increments between time steps. 

The gate mechanisms are constructed by a sigmoid neural network layer responsible for transforming a certain input into outputs with values in the range $[0, 1]$. These values are called keys, since they are used to tune the data passing through the gates, with 0 erasing the data, 1 passing it integrally and $]0, 1[$ passing it partially.

The first gate, also known as the forget gate, is responsible for controlling how much each information from the previous cell state $C_{t-1}$ will pass on to $C_t$. Its keys ($f_t$) are defined as
\begin{align}
    & f_t = \sigma (W_f \hspace{0.1cm} \cdot [\bar{y}_{t-1}, x_t] + b_f),
\end{align}
where $\sigma$ is the sigmoid activation function, $W_f$ is a weight matrix, $[\bar{y}_{t-1}, x_t]$ is the concatenation of the output from the previous cell with the input of the current cell and $b_f$ is a bias.

The second gate, known as the input gate, controls the intensity with which the information obtained from the current cell will be added to the new cell state. The new information candidate $\tilde{C}_t$ and the keys $i_t$ that regulate its intensity are defined as follows
\begin{align}
    & i_t = \sigma (W_i \hspace{0.1cm} \cdot [\bar{y}_{t-1}, x_t] + b_i) \\
    & \tilde{C}_t = tanh(W_C \hspace{0.1cm} \cdot [\bar{y}_{t-1}, x_t] + b_C),
\end{align}
with that, all the mechanisms necessary to define the new cell state $C_t$ were described. While the forget gate controls how much of the old cell state data will continue to pass, the input gate controls how much of the new information will be added to the cell state. To update it, the expression is defined as
\begin{align}
    & C_t = f_t \times C_{t-1} + i_t \times \tilde{C}_t.
\end{align}

The third and final gate, also called the output gate, is responsible for defining how much of the cell state information will be passed to the output $\bar{y}_t$ of the cell and the hidden state. Before using the information from the cell state, the hyperbolic tangent function is applied to reduce the range of its values. The keys ($o_t$) and the final output of the cell ($\bar{y}_t$) are given by
\begin{align}
    & o_t = \sigma(W_o [\bar{y}_{t-1}, x_t] + b_o), \\
    & \bar{y}_t = o_t \times tanh(C_t).
\end{align}

Now that the main features of the LSTM architecture and the motivation for using it have been highlighted, the next step is to organize the overall machine learning architecture that will be used to predict certain features.

The most interesting features to predict, whether in a numerical or in a experimental scenario, are divided in two categories depending on the output type: (i) static predictions, where each predicted feature has only one value, and (ii) transient predictions, where each predicted feature has its values changing through time. Our work focuses on the transient predictions, but an exploration of a static case is also provided in Section \ref{sec:dn}.

Both static and transient features can be used in the network input layer. Since the input dimensions need to be the same for all features, all static data is transformed into constant time series over all time steps. For example, when the dimensionless number $Re$ is used as an input, it is converted into a constant time series $Re(0) = Re(1) = \cdots  = Re(n_{max}) = Re$.

\begin{figure}[H]
    \centering
    \includegraphics[scale=0.4]{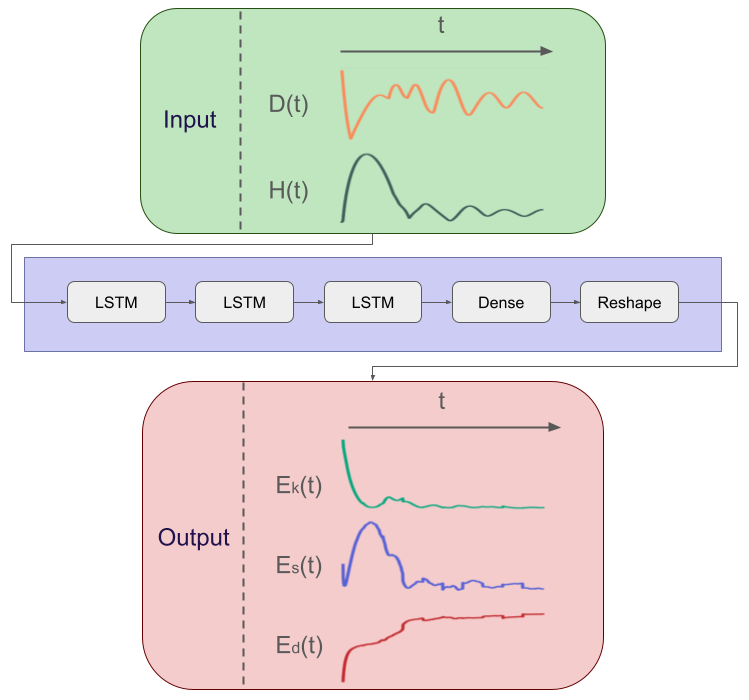}
    \caption{Machine Learning architecture for transient energy predictions using transient geometric data for the droplet collision problem. }
    \label{fig:organizacao_2}
\end{figure}

The architecture consists of LSTM hidden layers, followed by a Dense layer to generate values within the expected range for predictions, and a Reshape layer to adjust the output shape. The schematics for the input/output relations of one of the explored neural network structures for the droplet collision problem is shown in Fig. \ref{fig:organizacao_2}.

Table \ref{tab:testes} summarizes the test cases, characterized by output type (static or transient), dataset used, input features and output features.

\begin{table}[H]
    \begin{center}
        \begin{tabular}{ |p{1.75cm}||p{1.75cm}|p{3.5cm}|p{3cm}| }
            \hline
            \multicolumn{4}{|c|}{Test cases} \\
            \hline
                Type & Dataset & Input & Output \\
            \hline
                Transient & Spreading & $t, D(t)$, Re, We & $E_k(t), E_s(t), E_d(t)$ \\
                Transient & Collision & $D(t), H(t)$, Re, We, B & $E_k(t), E_s(t), E_d(t)$ \\
                Transient & Collision & $D(t), H(t)$ & $E_k(t), E_s(t), E_d(t)$ \\
                Static & Collision & $E_k(t), E_s(t), E_d(t)$ & Re, We \\
            \hline
        \end{tabular}
        
        \caption{Summary of the explored test cases.}
        \label{tab:testes}
    \end{center}
\end{table}

\section{Results}\label{sec:results}

A supervised LSTM network was employed to train models capable of predicting the energy budget based on geometric time series data extracted from numerical simulations. Initially, to test the predictive performance and adaptability of the architecture, two models were trained by using each of the previously defined datasets. 

\subsection{Performance of the model formulations}

Firstly, as the hyperparameter tuning is a crucial stage for applying machine learning modeling, in the current work we have employed the open source hyperparameter optimization framework \textit{Optuna} \cite{Optuna2024}. Following a detailed analysis of datasets from simulations of droplets with varying initial shapes impacting on solid surfaces and droplet collisions, the hyperparameters were chosen as follows

\begin{itemize}
    \item Learning rate of 0.0001;
    \item Maximum of 5000 epochs;
    \item Batch sizes of 8 for spreading and 32 for collision;
    \item 3 LSTM hidden layers with 32 neurons each;
    \item 1 Dense output layer with $n \times m$ neurons, where $n$ represents the number of time steps and $m$ represents the number of features;
    \item 1 Reshape layer (transient output only);
    \item Adam as an optimization algorithm;
    \item Mean square error applied as the loss function.
\end{itemize}

Additionally, an early stopping is applied based on training loss (collision) or validation loss (spreading).

To measure the predictive capability of each model, we used the Coefficient of determination - $R^2$ (\ref{eq:r2}), the Root Mean Square Error - $RMSE$ (\ref{eq:rmse}) and the Normalized Root Mean Square Error - $NRMSE$ (\ref{eq:nrmse}) defined, respectively, as

\begin{align}
    & R^2 = 1 - \frac{\sum_{i=1}^{N} (y_i - \tilde{y}_i)^2}{\sum_{i=1}^{N} (y_i - \bar{y})^2}, \label{eq:r2} \\
    & RMSE = \sqrt{\frac{1}{N} \sum_{i=1}^{N}(y_i - \tilde{y}_i)^2}, \label{eq:rmse} \\
    & NRMSE = \frac{RMSE}{y_{max} - y_{min}}, \label{eq:nrmse}
\end{align}
where $N$ is the number of samples, $y_{max}$ and $y_{min}$ are, respectively, the maximum and the minimum possible value for the feature, $y_i$ and $\tilde{y}_i$ are, respectively, the expected and the predicted value, and $\bar{y}$ is the average of all expected values. 

The $R^2$ is employed to assess how well the variability of the dependent variables (energies) can be explained by the variability of the independent variables (geometric data and dimensionless numbers). A score approaching 0 indicates a poor explanatory capability, while a score nearing 1 signifies a strong explanatory capability for the dependent variable. $RMSE$ and $NRMSE$ are both used to evaluate the quality of the predictions. While $RMSE$ generates a value dependent on the scale of the predicted variables, $NRMSE$ is used to provide a relative measure of error regardless of the data scale. In both cases, a value close to zero indicates a well-fitted model.

While metrics (\ref{eq:r2}) and (\ref{eq:rmse}) are used to individually measure the efficiency of each model, (\ref{eq:nrmse}) is used to compare models and check the adaptability of the chosen architecture.

For the spreading dataset, 10 different shapes are split into 70\% training (7 shapes, 95 samples), 20\% validation (2 shapes, 27 samples) and 10\% test (1 shape, 10 samples). To optimize the training process, validation loss is used as an early stopping control parameter. 

The $R^2$ score for each individual sample ranges from 0.98315 to 0.9994, while the $RMSE$ ranges from 0.01122 to 0.05893. Overall, both metrics fall within the expected optimal range and the majority of the predictions fall within a $\pm 10\%$ error range. Additionally, we observe that these metrics remain rather consistent across all samples, providing further evidence that the model is well-fitted.

For the collision dataset, 72 samples are split into 70\% training (50 samples) and 30\% test (22 samples). To optimize the training process, training loss is used as an early stopping control parameter.

The $R^2$ score for each of the 22 test samples ranges from 0.95291 to 0.99997, with the $RMSE$ ranging from 0.00115 to 0.04618. Given the increased variability in energy behavior, it is understandable how the range of the metrics is larger than in the spreading case. Another contributing factor is that there are less samples in the training dataset with higher Reynolds values, resulting in lower accuracy for predictions within that range.

Despite these differences, both metrics remain near optimal ranges, even presenting higher accuracy with fewer training samples than the spreading dataset, as will be shown next.

Previously, the metrics were analyzed for each sample individually. To assess the overall prediction accuracy of the models, Table \ref{tab:metricas} compares the error metrics for all training and test samples from both datasets.

\renewcommand{\arraystretch}{1.5}
\begin{table}[H]
    \begin{center}
        \begin{tabular}{ |p{2cm}||p{2cm}|p{2cm}|p{2cm}|p{2cm}| }
            \hline
                \multirow{2}{*}{Metric} & \multicolumn{2}{c|}{Training samples} & \multicolumn{2}{c|}{Test samples} \\
            \cline{2-5}
                & \multicolumn{1}{c|}{Spreading} & \multicolumn{1}{c|}{Collision} & \multicolumn{1}{c|}{Spreading} & \multicolumn{1}{c|}{Collision} \\
            \hline
                $R^2$ & 0.9924 & 0.99996 & 0.99429 & 0.99559 \\
                $RMSE$ & 0.03677 & 0.00124 & 0.02957 & 0.0137 \\
                $NRMSE$ & 0.03674 & 0.0019 & 0.02961 & 0.02096 \\
            \hline
        \end{tabular}
        \caption{Comparison of the evaluation metrics for each dataset.}
        \label{tab:metricas}
    \end{center}
\end{table}
\renewcommand{\arraystretch}{1.0}

Since the training datasets are employed in training the model, it is expected that their metrics should also be optimal. Nevertheless, comparing these metrics with those of the test datasets provides insights into how well the model has acquired adaptability to new sets of data that were not used during training. 

The most important metric for comparison between both models is the $NRMSE$, as it offers a normalized measure of error, disregarding the scale of the data. With these insights, we can infer that the chosen architecture is not only reasonably accurate but also highly versatile for different regimes. 

To better emphasize the need for a more complex model such as the LSTM, we compared predictions generated by linear regression, a simple RNN, and our LSTM on the collision dataset.

\begin{figure}
\centering
    \subfigure[Linear regression]{
        \includegraphics[width=0.95\textwidth]{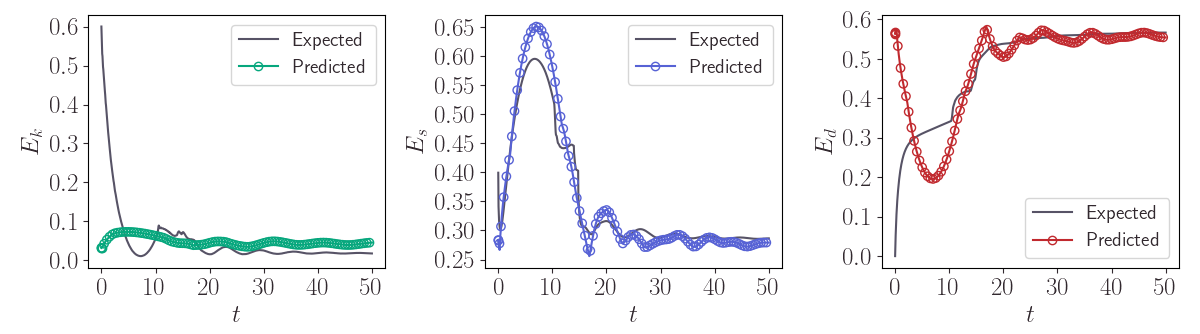}
        \label{fig:comparacao_regressao}
    }
    \subfigure[RNN]{
        \includegraphics[width=0.95\textwidth]{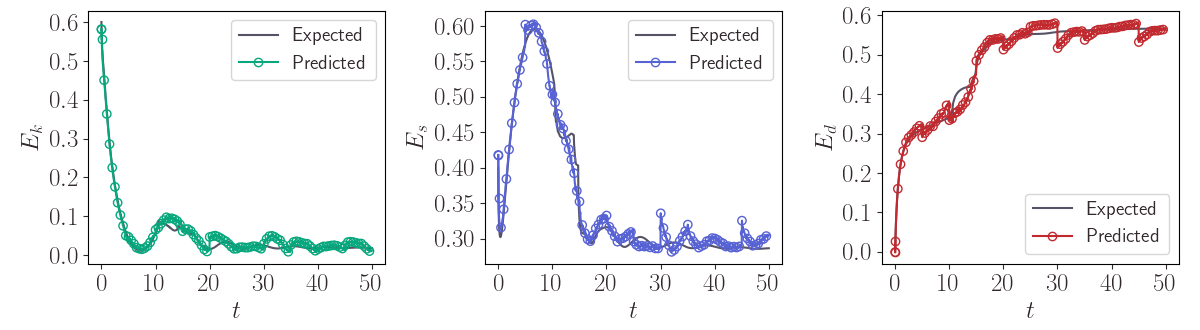}
        \label{fig:comparacao_rnn}
    }
    \subfigure[LSTM]{
        \includegraphics[width=0.95\textwidth]{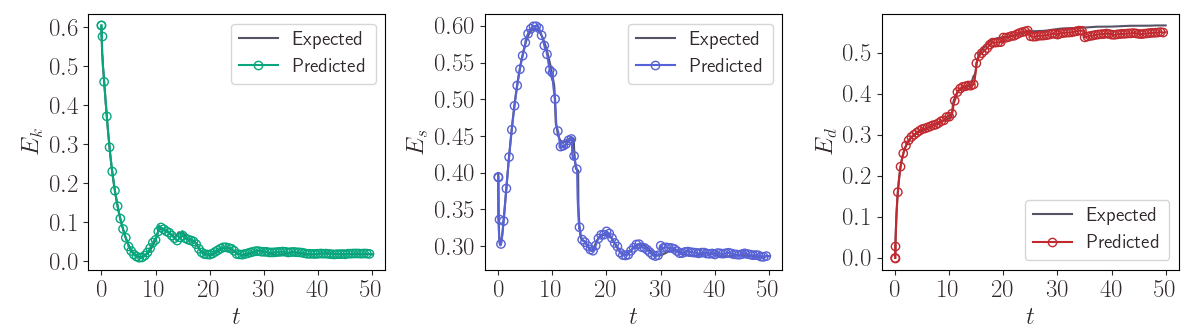}
        \label{fig:comparacao_lstm}
    }
    \caption{Comparison of the predictions given by different machine learning models. All cases correspond to a binary droplet collision with Re = 115.56, We = 53.0, B = 0.17.}
    \label{fig:comparacao_modelos}
\end{figure}

As shown in the metrics in Table \ref{tab:metricas_modelos} and the prediction example in Fig. \ref{fig:comparacao_modelos}, the LSTM model (Fig. \ref{fig:comparacao_lstm}) is essential for accurately capturing the complex behavior of the analyzed dynamics. While linear regression (Fig. \ref{fig:comparacao_regressao}) fails to capture the temporal dynamics of this problem, the RNN (Fig. \ref{fig:comparacao_rnn}) partially succeeds but still struggles to maintain context over many time steps.

\renewcommand{\arraystretch}{1.5}
\begin{table}[H]
    \begin{center}
        \begin{tabular}{ |p{2cm}||p{2cm}|p{2cm}|p{2cm}|p{2cm}| }
            \hline
                Metrics & \multicolumn{1}{c|}{Linear regression} & \multicolumn{1}{c|}{RNN} & \multicolumn{1}{c|}{LSTM} \\
            \hline
                $R^2$ & 0.9201 & 0.9652 & 0.99559 \\
                $RMSE$ & 0.0583 & 0.0385 & 0.0137 \\
                $NRMSE$ & 0.0892 & 0.0589 & 0.02096 \\
            \hline
        \end{tabular}
        \caption{Comparison of the evaluation metrics for each technique.}
        \label{tab:metricas_modelos}
    \end{center}
\end{table}
\renewcommand{\arraystretch}{1.0}

Finally, another important aspect is the computational feasibility of the method. To assess the difference in computational cost between performing a full numerical simulation up to $t_{\text{max}} = 50$ for each parameter and obtaining energy predictions through a trained model, time evaluation tests were conducted on similar hardware. Each numerical simulation, for both the spreading and collision cases, requires approximately 8 to 12 hours, depending on the complexity of the simulated dynamics. The offline training phase of all models takes 3 to 6 hours, depending on the early stopping tolerance, while the online testing phase takes, on average, 0.0729s to predict all 1000 time steps up to $t_{\text{max}} = 50$ for each parameter set.

\subsection{Prediction for the energy budget}

\begin{figure}
\centering
    \subfigure[Re = 30.73, We = 1659.38.]{
        \includegraphics[scale=0.51]{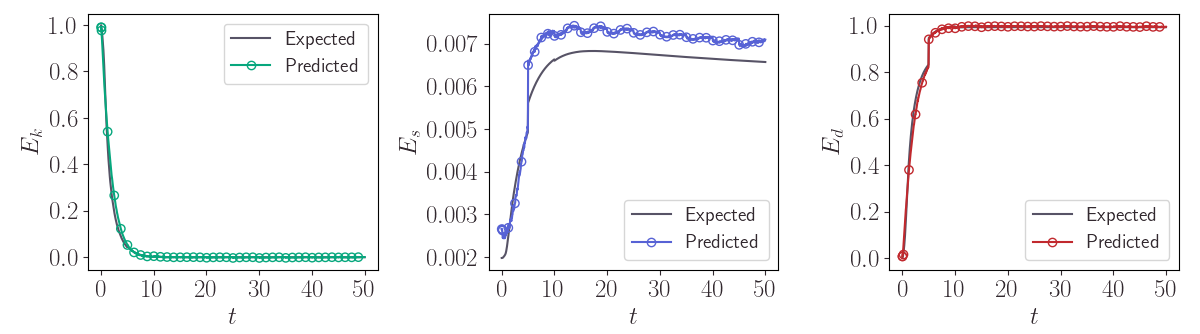}
        \label{fig:predictEt_aprox_shape_1}
    }
    \subfigure[Re = 24.81, We = 90.62.]{
        \includegraphics[scale=0.51]{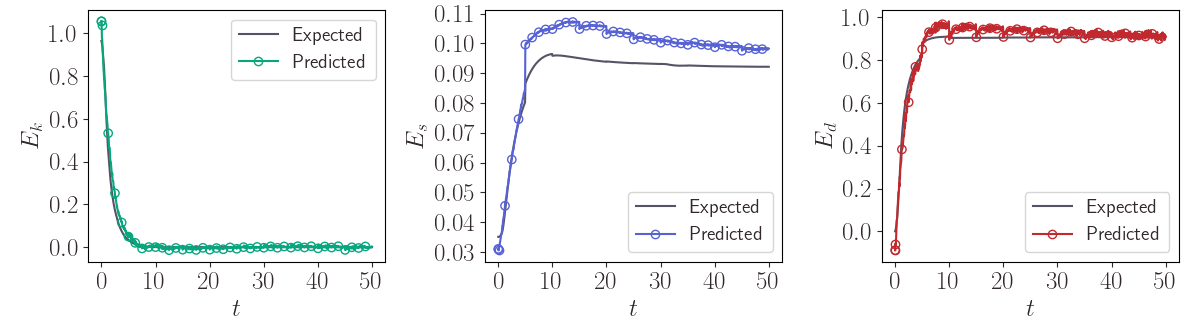}
        \label{fig:predictEt_aprox_shape_2}
    }
    \caption{Comparison between the expected and predicted energy time series from two samples of the spreading test dataset.}
    \label{fig:comparacao_shape}
\end{figure}

To better visualize the model's efficiency for the spreading case, we first analyze the prediction behavior of two different samples shown in Fig. \ref{fig:comparacao_shape}. Each energy plot is normalized by the total energy in the first time step $E_t(0)$.

In Figs. \ref{fig:predictEt_aprox_shape_1} and  \ref{fig:predictEt_aprox_shape_2}, we can observe pronounced fluctuations caused by the sliding technique (as depicted in Fig. \ref{fig:organizacao_1}) applied to the dataset. Given that each sample is divided into multiple sub-samples, the predictions can oscillate between each one of them, as observed in the surface energy $E_s$ shown in Fig. \ref{fig:predictEt_aprox_shape_1} and the viscous dissipation $E_d$ from Fig. \ref{fig:predictEt_aprox_shape_2}. While it may appear to increase errors, some tests have shown that the technique, in fact, enhances model accuracy by generating more samples and using fewer time steps, which reduces training complexity.

Since most samples from the training and test datasets exhibit a lower range of surface energy ($E_s$) values, as illustrated in Fig. \ref{fig:predictEt_aprox_shape_1}, their prediction errors end up being the most visually accentuated. 

The predominance of samples within this lower range also influences the training process. Samples with lower Weber numbers, which have higher surface tension effects, exhibit higher surface energies compared to the majority. As depicted in Fig. \ref{fig:predictEt_aprox_shape_2}, this higher surface energy undermines accuracy, as the model is primarily tuned for lower values. 

In general, the main difference in energy behavior between different entries in the spreading dataset lies in the steepness of their curves (see Fig. \ref{fig:conjunto1_1}). Since the initial drop shape is one of the key factors influencing this behavior, the analysis in \ref{fig:comparacao_shape} indicates that the model is capable of generating accurate predictions for samples with a new drop shape. 

\begin{figure}
\centering
    \subfigure[Re = 115.56, We = 53.0, B = 0.17.]{
        \includegraphics[scale=0.51]{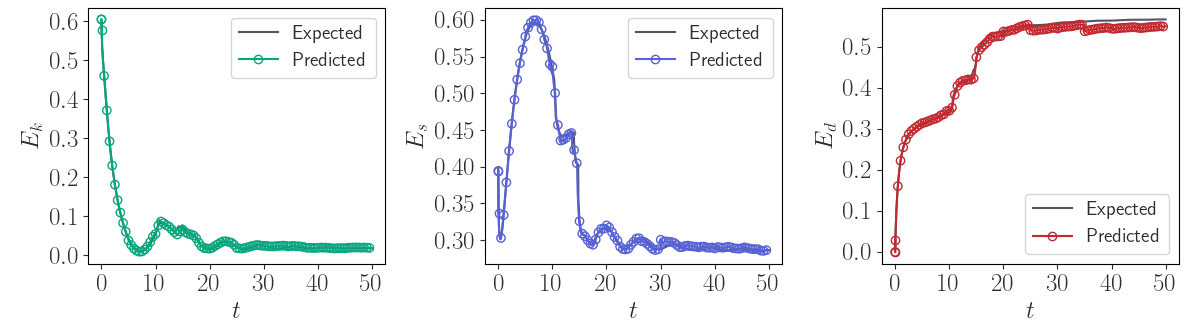}
        \label{fig:predictEt_aprox_oscilacao_1}
    }
    \subfigure[Re = 74.45, We = 22.0, B = 0.27.]{
        \includegraphics[scale=0.51]{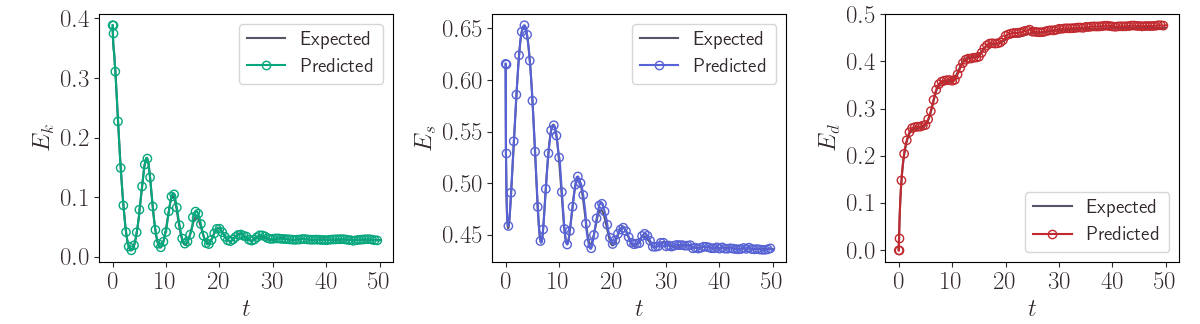}
        \label{fig:predictEt_aprox_oscilacao_2}
    }
    \subfigure[Re = 120.45, We = 58.0, B = 0.17.]{
        \includegraphics[scale=0.51]{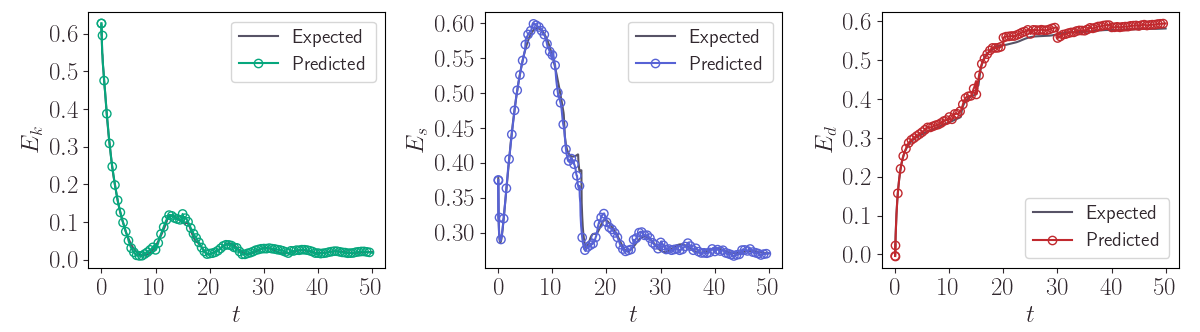}
        \label{fig:predictEt_aprox_oscilacao_3}
    }
    \caption{Comparison between the expected and predicted energy time series from three samples of the collision test dataset. Subfigures (a) and (b) correspond to interpolation cases, with Re values within the training range of [74.45, 115.56], while subfigure (c) represents an extrapolation case with Re = 120.45.}
    \label{fig:comparacao_oscilacao}
\end{figure}

For the binary collision dataset, we illustrate in Fig. \ref{fig:comparacao_oscilacao} the prediction behavior of three different test cases. Again, the energy plot is normalized using the total energy in the first time step.

The collision adds another layer of complexity to the problem, as the energies now exhibit oscillations over time. In Fig. \ref{fig:predictEt_aprox_oscilacao_1} we observe that the model can produce accurate predictions even when there is an anomaly in the curves (as seen at time $t \approx 6$). Furthermore, although fluctuations in viscous dissipation $E_d$ are still detectable, they occur much less frequently. Decreasing Re and We, as illustrated in Fig. \ref{fig:predictEt_aprox_oscilacao_2}, the coalesced droplet experiences an oscillatory regime that can also be correctly captured by the prediction. Overall, as shown in these figures, we can confirm the excellent agreement between the expected and predicted results in the interpolation scenarios.

All samples corresponding to $Re=120.45$ were reserved for the test set to evaluate the model's performance in extrapolation scenarios. Fig. \ref{fig:predictEt_aprox_oscilacao_3} demonstrates that the model still provides a reasonable qualitative prediction for the extrapolated value.

\begin{figure}
\centering
    \subfigure[Re = 74.45, We = 22.0, B = 0.19.]{
        \includegraphics[width=0.45\textwidth]{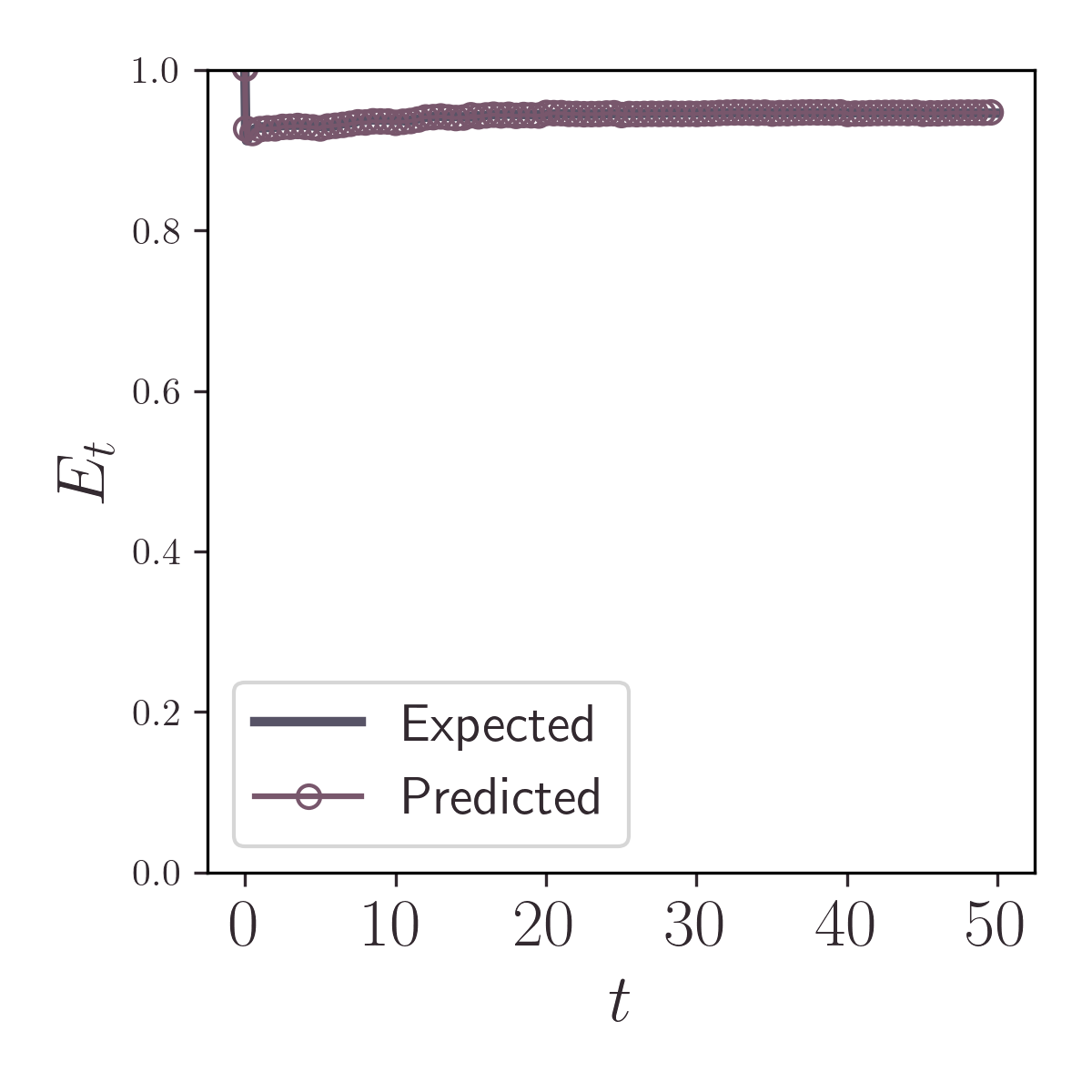}
        \label{fig:energia_total1}
    }
    \subfigure[Re = 85.48, We = 29.0, B = 0.19.]{
        \includegraphics[width=0.45\textwidth]{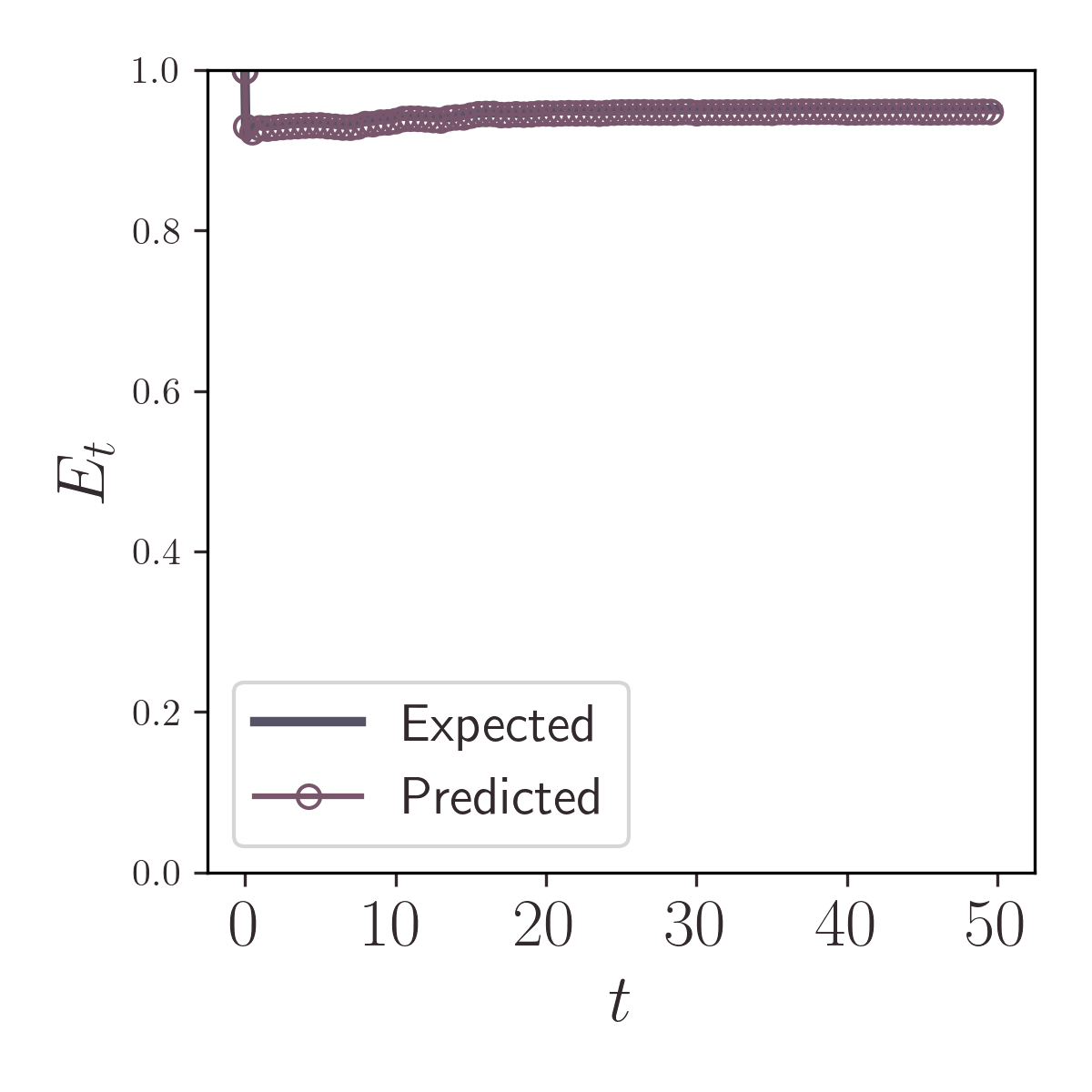}
        \label{fig:energia_total2}
    }
    \caption{Examples of total energy behavior for different parameter sets.}
    \label{fig:energia_total}
\end{figure}

To ensure the robustness of our machine learning strategy in addressing a physical constraint, we have included results concerning the total energy conservation. 
To illustrate this, Fig. \ref{fig:energia_total} presents a comparison between the predicted and expected behavior of the total energy for different parameters. Due to numerical errors in the training data, which arise from using an intermediate mesh refinement, some slight fluctuations in energy over time can be observed. As the learning algorithm attempts to model these dynamics, the generated forecasts tend to replicate this fluctuating pattern, but mostly maintain the total energy. It is important to highlight that according to our tests, no physically impossible cases were observed in the predictions.

\subsection{Two-phase prediction with sequential neural networks}

To further investigate the versatility of our proposed methodology, we adopt a two-phase prediction strategy tailored specifically for the droplet collision problem. Initially, we consider only the droplet diameter $D(t)$ and height $H(t)$ 
as inputs to our model. In this initial phase, our aim is to accurately estimate the energy budget associated with the collision process. Subsequently, leveraging the insights gained from the energy budget prediction, we extend our model to predict static values, such as the nondimensional parameters $Re$ and $We$. 

This two-phase prediction approach is particularly valuable in experimental contexts where precise fluid parameters measurements are difficult to obtain. Often, despite having access to video data from experiments, the lack of comprehensive knowledge regarding the fluid properties impedes accurate prediction of the energy budget. In such scenarios, our method provides a robust solution by including geometric data to not only predict the energy budget but also estimate dimensionless parameters essential for understanding the underlying physics of droplet dynamics. Using this innovative two-phase prediction process, we bridge the gap between experimental observations and theoretical insights, thereby enhancing our ability to analyze and interpret droplet collision phenomena with greater precision and efficiency.

\begin{figure}
    \centering
    \includegraphics[scale=0.45]{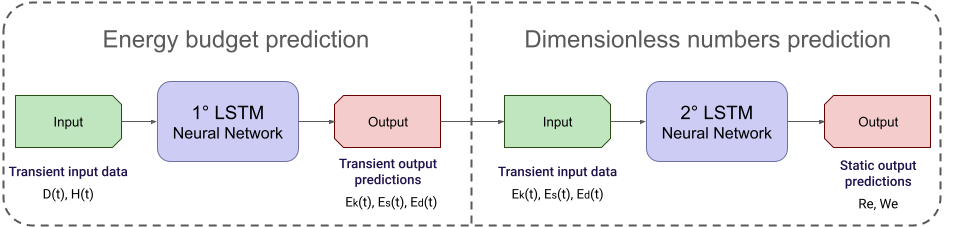}
    \caption{Two-phase schematics for the energies and dimensionless numbers predictions.}
    \label{fig:organizacao_3}
\end{figure}

In summary, this process involves two separate neural networks. The output of the first network, which predicts the energy budget, serves as input for predicting the dimensionless numbers in the second network, as illustrated in Fig. \ref{fig:organizacao_3}.

\subsubsection{Energy budget prediction}

\begin{figure}
\centering
    \subfigure[With dimensionless numbers.]{
        \includegraphics[scale=0.51]{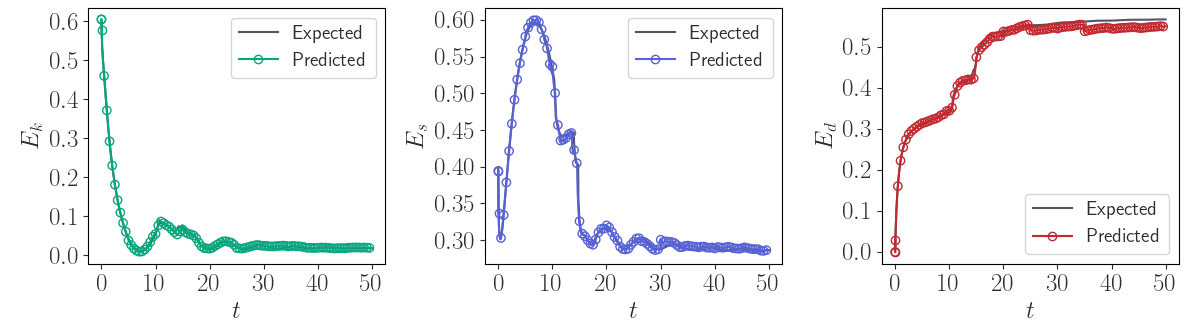}
        \label{fig:predictEt_aprox_oscilacao_comparacao_1}
    }
    \subfigure[Without dimensionless numbers.]{
        \includegraphics[scale=0.51]{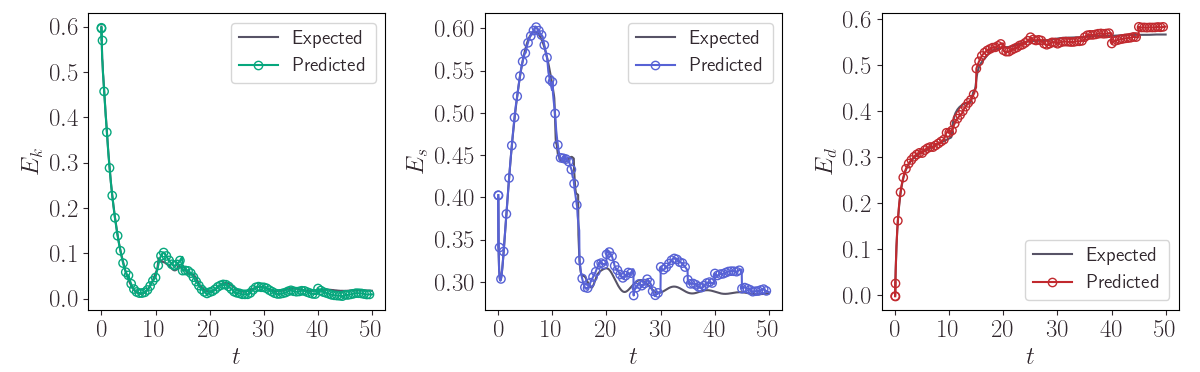}
        \label{fig:predictEt_aprox_oscilacao_comparacao_2}
    }
    \caption{Comparison between energy predictions with and without the dimensionless numbers as inputs - Re = 115.56, We = 53.0, B = 0.17.}
    \label{fig:predictEt_aprox_oscilacao_comparacao}
\end{figure}

For the first step, we employ an LSTM network to predict all three energies using only the diameter $D(t)$ and height $H(t)$ of the droplet. Fig. \ref{fig:predictEt_aprox_oscilacao_comparacao} illustrates an example of how the energy prediction behavior is affected when dimensionless numbers are omitted as inputs.

While omitting the dimensionless numbers reduces the model's accuracy, it still effectively captures the overall behavior of the energies. Moreover, given the high difficulty of measuring energy data from experimental videos through traditional methods, this increase in error remains tolerable. The differences in metrics across training and test samples for both configurations can be seen in Table \ref{tab:metricas_input}.

\renewcommand{\arraystretch}{1.5}
\begin{table}[H]
    \begin{center}
        \begin{tabular}{ |p{2cm}||p{2cm}|p{2cm}|p{2cm}|p{2cm}| }
            \hline
                \multirow{2}{*}{Metric} & \multicolumn{2}{c|}{Training samples} & \multicolumn{2}{c|}{Test samples} \\
            \cline{2-5}
                & \multicolumn{1}{c|}{With} & \multicolumn{1}{c|}{Without} & \multicolumn{1}{c|}{With} & \multicolumn{1}{c|}{Without} \\
            \hline
                $R^2$ & 0.99996 & 0.99963 & 0.99604 & 0.9776 \\
                $RMSE$ & 0.00124 & 0.00401 & 0.02732 & 0.03087 \\
                $NRMSE$ & 0.0019 & 0.00615 & 0.02736 & 0.04725 \\
            \hline
        \end{tabular}
        \caption{Comparison of the evaluation metrics for models with and without dimensionless numbers Re, We and B as inputs.}
        \label{tab:metricas_input}
    \end{center}
\end{table}
\renewcommand{\arraystretch}{1.0}

In general, if the dimensionless numbers are known, it is advisable to choose the model that has been trained with them included. However, upon comparing the metrics, it is evident that utilizing only geometric data for the model remains a viable option with decent accuracy.

\subsubsection{Dimensionless numbers prediction} \label{sec:dn}

In a parallel test, we attempted to predict the dimensionless numbers directly using only geometric data. However, this approach resulted in poor prediction accuracy for the collision dataset, likely due to the weak correlation between these terms, as discussed in Appendix \ref{ap:direto}. 

Therefore, rather than directly using $D(t)$ and $H(t)$ to predict the dimensionless numbers $Re$ and $We$, the second step involves first utilizing the previously trained neural network to predict the energies, then training a new neural network to predict the dimensionless numbers based on these energies.

\begin{figure}[H]
\centering
\includegraphics[scale=0.55]{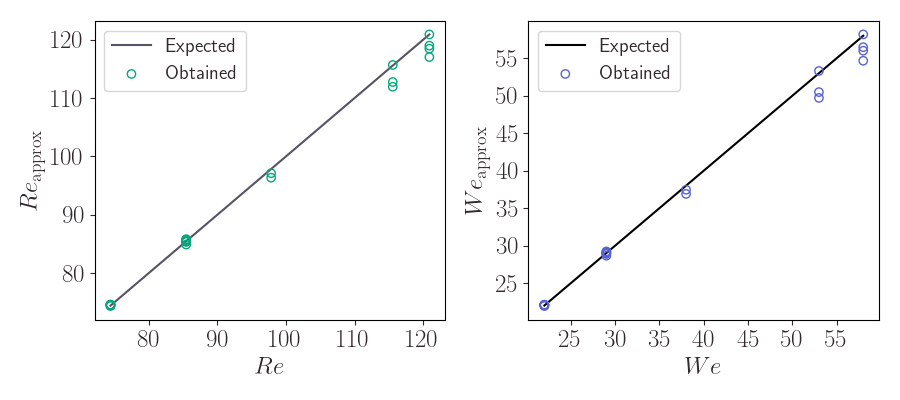}
\caption{Dimensionless numbers predictions for each of the 22 samples from the droplet collision test dataset.}
\label{fig:predictReWe_aprox}
\end{figure}

Given that each sample is divided into 100 time step slices, the selected prediction for a complete sample (consisting of 1000 time steps) is determined by taking the median of the predicted values from its 10 slices. This is done to mitigate potential noise caused by high errors resulting from using approximated energies as inputs. The prediction results for the collision test dataset are depicted in Fig. \ref{fig:predictReWe_aprox}.

Despite the introduction of some noise through the use of approximated energies, the overall predicted values still appear to align with the expected. To quantitatively analyze the results, the previous metrics are calculated both for training and test data in Table \ref{tab:metricas_numeros}.

\renewcommand{\arraystretch}{1.5}
\begin{table}[H]
    \begin{center}
        \begin{tabular}{ |p{2cm}||p{4cm}|p{4cm}| }
            \hline
                Metric & Training samples & Test samples \\
            \hline
                $R^2$ & 0.99366  & 0.99823 \\
                $RMSE$ & 2.50602 & 1.39455 \\
                $NRMSE$ & 0.02534 & 0.0141 \\
            \hline
        \end{tabular}
        \caption{Comparison of the evaluation metrics for the prediction of the dimensionless numbers by using predicted energies.}
        \label{tab:metricas_numeros}
    \end{center}
\end{table}
\renewcommand{\arraystretch}{1.0}

With this, we have demonstrated that, in the explored regimes, it is feasible to predict energies and dimensionless numbers using only geometric data. This input data is easily accessible from experimental videos, so that this approach can also be used in experiments. As a follow-up to this study, new architectures could be developed and tested with additional datasets to enhance the neural network generalization, thereby expanding its applicability to a wider range of regimes.

\section{Conclusions}

In this study, we proposed an LSTM-based model to predict the energy balance in two droplet dynamics problems: the impact of a non-spherical droplet on a rigid surface and the collision of two equal-sized droplets. For the first case, we also implemented a method to extract a Front-Tracking interface from experimental images allowing the use of realistic non-spherical droplet shapes in generating training data (see Appendix \ref{ap:extraction}). Following this test, a two-phase sequential prediction approach was applied to predict important properties using exclusively geometric data. The current methodology provides an easy-to-use tool for the challenging task of predicting important flow quantities such as the energy budget and the dimensionless numbers, which are useful for various industrial applications. As an example, optimizing atomizers or sprays for coating applications involves predicting whether the kinetic energy will be sufficient for the droplets to cover the entire surface and spread uniformly, and controlling the final droplet shape after impact, determining whether it will spread or maintain a more spherical form, through surface energy \cite{Nasser2011}. Additionally, in agriculture, analyzing surface energy as a metric to understand droplet interactions with surfaces, such as their behavior on hydrophobic leaves, could help improve the control of agricultural sprayers, while evaluating viscous dissipation to assess energy loss of droplets upon impact may also contribute to optimizing sprayer design \cite{Athira2024}. Finally, the food industry can also be mentioned, with the goal of modeling droplet behavior and controlling phenomena such as emulsion formation and droplet stability \cite{Andrade2013}.

For the one-phase predictions, both proposed models share the same overall architecture, differing only in the data used for training. The first model was trained using data from 95 numerical simulations with seven different shapes extracted from real experimental images of droplets impacting on solid surfaces. In this model, the neural network input consists of simulation time, the temporal series of the droplet diameter and nondimensional numbers. The second model utilized data from 50 numerical simulations of two colliding droplets that elongate and retract after coalescence, leading to oscillations in the energy budget. Due to the more complex oscillatory behavior, this model is initialized with the temporal series of diameter and height of the coalesced droplet, in combination with nondimensional numbers and the impact parameter. In both models, the output is the energy budget over time.

To evaluate the performance of each model, we conducted a comprehensive metric analysis commonly employed in machine learning modeling, which includes the Coefficient of Determination ($R^2$), the Root Mean Square Error ($RMSE$), and the Normalized Root Mean Square Error ($NRMSE$). The $R^2$, $RMSE$ and $NRMSE$ metrics for both models are, respectively, 0.99429 and 0.99559, 0.02957 and 0.0137, and 0.02961 and 0.02096. After analyzing the metrics, we observe that both models achieved performances sufficiently close to the ideal values for each metric, with $R^2$ close to one and $RMSE$ approaching zero. The analyzed models are accurate in a similar way, even when applied to considerably different problems, which can be evaluated by the similarity of their $NRMSEs$. The spreading model showed promising results by successfully handling a completely new shape. Similarly, the model for binary droplet collision performed well with samples featuring sets of parameters not used during training. The accuracy of the models is also evaluated through plots depicting the kinetic, surface, and viscous dissipation, confirming excellent agreement between the simulation data and the predicted values. This suggests that the machine learning architectures have the potential to be highly adaptable and accurate across a range of fluid regimes.

For the prediction of nondimensional numbers, the model architecture remained similar but followed a two-phase scheme. Firstly, an LSTM network predicts the energy budget using only geometric data. Then, the predicted energies serve as inputs in a second LSTM network to predict the static dimensionless numbers. The metrics for the first phase (energies) and the second phase (static prediction for Reynolds and Weber numbers) are, respectively, 0.9776 and 0.99823, 0.03087 and 1.39455, and 0.04725 and 0.0141. Despite a slight increase in error when compared to the one-phase prediction strategy, the metrics for two-phase predictions still demonstrate high accuracy. Results show excellent agreement for low values of Re and We, whereas the neural network has more difficulty in predicting high values of $Re$ and $We$. This difficulty can be attributed to the fact that, for the training data, there are fewer instances with higher Re and We values, as seen in Table \ref{tab:conjunto2}. Finally, this approach proves useful in the sense that directly attempting to predict the dimensionless numbers yielded higher errors, as demonstrated in Appendix \ref{ap:direto}.

While the metrics indicate high versatility, some questions must still be explored. One notable challenge is the prediction of regimes beyond coalescence in the binary droplet collision scenario, as transitioning between regimes presents a complex task for machine learning models. This difficulty is further accentuated in our model due to its reliance on geometric data ($D(t)$ and $H(t)$) as inputs, which are significantly affected by regimes with bouncing or breakup. Future research could focus on the interpretation of the model itself, as well as modifications to the LSTM architecture, such as the Bidirectional LSTM \cite{Sima2019}, Attention LSTM \cite{Xianyun2023}, or adaptations to make it physics-informed \cite{Dong2024}.

We hope that this research will shed light on the advantages of leveraging the performance offered by LSTM architectures for predicting different types of transient data in complex flows. Although our current work does not address complex fluids, it establishes a foundation for future studies that could extend these methods to address more challenging scenarios, such as the dynamics of viscoelastic droplets \cite{Oishi2019, Hugo2022_2}.

\section*{Acknowledgements}
The authors would also like to thank the financial support given by the São Paulo Research Foundation (FAPESP) grants \#2013/07375-0 and \#2021/14953-6, the National Council for Scientific and Technological Development (CNPq), grants \#307228/2023-1
and CAPES. The authors also acknowledge the Numerical Simulation and AI Laboratory at FCT/UNESP for their support with cluster resources. C.M. Oishi would like to thank the insightful discussions on machine learning modeling held with the research team during his visit at the AI Institute in Dynamic Systems, University of Washington.

\section*{Data Availability Statement}
The algorithm employed to generate the dataset, along with the machine learning codes supporting the findings of this study, are available upon reasonable request from the corresponding author.

\begin{appendices}

\section{Drop shape extraction from an image} \label{ap:extraction}

In experimental environments, working with perfectly spherical or elliptical droplets is practically an impossible task. However, when reproducing these experiments through numerical simulations, spherical droplets are often used as an approximation due to their simple definitions through parametric functions.

To avoid using spherical approximations, a method capable of extracting Front-Tracking interfaces from experimental images was developed in this work. To extract particles from an image, the steps shown in Fig. \ref{fig:extracao_1} are performed. In more detail, these steps are:

\begin{figure}[H]
\centering
\includegraphics[scale=0.35]{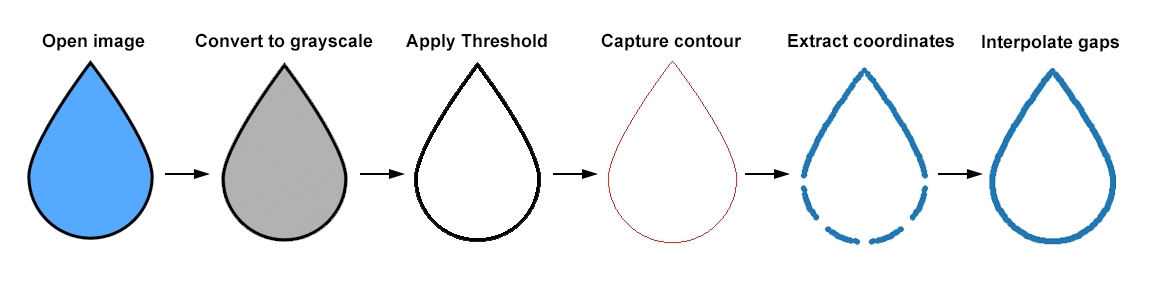}
\caption{Steps to extract particles from an image.}
\label{fig:extracao_1}
\end{figure}

\begin{description}
\item[Step 1:] open an image that contains the interface. Preferably, the interface should be outlined with another color to help the contour detection process;

\item[Step 2:] change the image to grayscale or directly binarize to black and white (recommended only for cases that feature easily extractable interfaces with minimal noise);

\item[Step 3:] binarize the grayscale image through thresholding. Pixels are binarized depending on if their values are above or below a given threshold parameter. In Fig. \ref{fig:extracao_1}, a threshold of $L = 110$ was applied;

\item[Step 4:] extract the contour(s) of the image that represents the interface(s). In the example previously shown, there is only one continuous interface, so only one contour was detected;

\item[Step 5:] approximate the contour shape by collecting a set of discretized points $(i, j)$. Since the points collected by the function are actually the indices of the pixels from the image contour, they need to be adjusted so that they fall within the ranges $[x_{min}, x_{max}]$ and $[y_{min}, y_{max}]$:

\begin{align}
& x = \frac{j(x_{min}-x_{max})}{m} + x_{max}, \\
& y = \frac{i(y_{min}-y_{max})}{n} + y_{max},
\end{align}

With $n \times m$ being the dimensions of the input image and $(x, y)$ being the new coordinates for the points extracted from the interface. As the list of pixels returned is ordered, the method facilitates the creation of particles for representation by front-tracking.

\item[Step 6:] If the function returns an insufficient number of points or generates points with large spacing (as seen in the previous example), it is possible to apply linear interpolation to create new intermediate points.

To do this, the Euclidean distance between each pair of points is calculated and, if the distance is greater than a chosen maximum tolerance, new intermediate points are generated.

\end{description}

\begin{figure}
\centering
    \subfigure[Aliasing generated when extracting points from pixels.]{
        \includegraphics[scale=0.4]{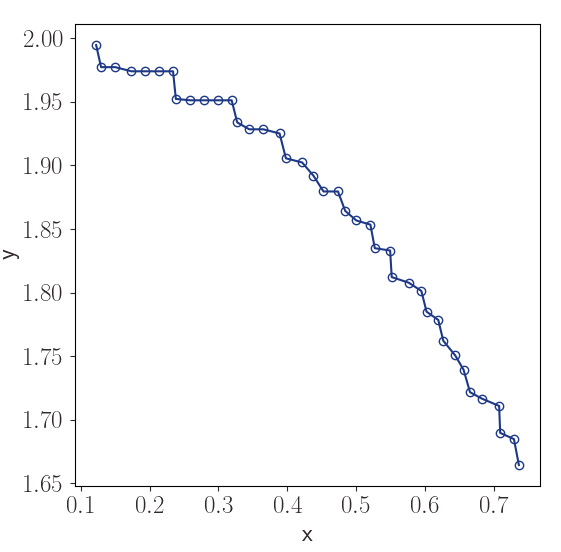}
        \label{fig:solucao1_1}
    }
    \subfigure[The new points after applying TSUR to reduce aliasing.]{
        \includegraphics[scale=0.4]{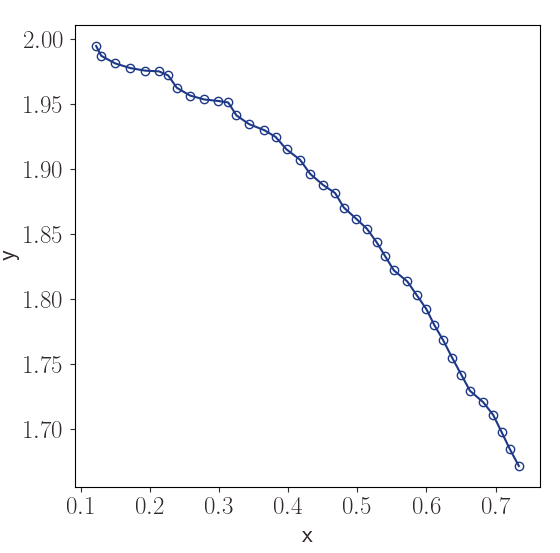}
        \label{fig:solucao1_2}
    }
    \caption{Comparison before and after applying the TSUR technique.}
    \label{fig:solucao_comparacao}
\end{figure}

As the method involves extracting points through the pixelated image contour, the main limitation to this approach is the image resolution. Additionally, as the points are extracted from the lower left corner of each pixel, aliasing behavior is generated in the extracted points, as can be seen in Fig. \ref{fig:solucao1_1}. This disturbance in the initialization of points can generate atypical behavior in the earlier steps of the simulation.

To help solve the aliasing problem, a technique was applied directly to the solver. The Trapezoidal Sub-grid Undulations Removal (TSUR), proposed in \cite{Nonato2004}, is a mass conserving smoothing method used to reduce aliasing and abrupt undulations in curves. The result of applying the method is shown in Fig. \ref{fig:solucao1_2}. 

In addition to the effectiveness of TSUR in removing the initial aliasing of the interface, the technique can also reduce numerical noise that may appear during the simulation, generating more physically accurate Front-Tracking solutions.

\section{Static predictions for the dimensionless numbers using geometric data} \label{ap:direto}

\begin{figure}[H]
\centering
\includegraphics[scale=0.55]{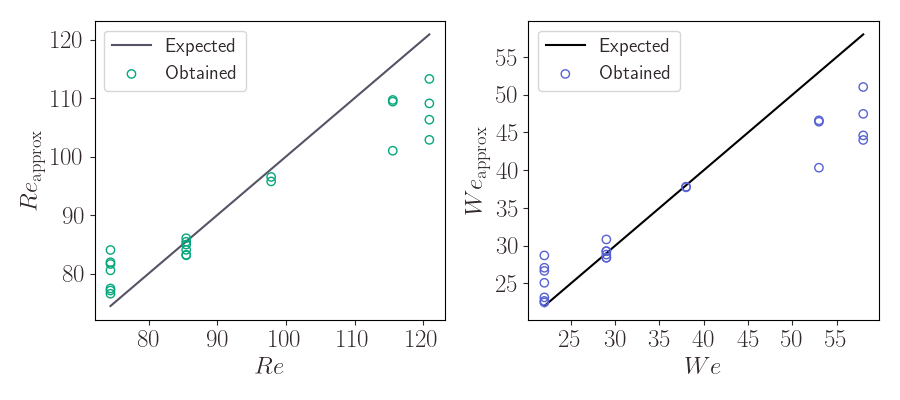}
\caption{Direct dimensionless numbers predictions for each of the 22 samples from the collision test dataset.}
\label{fig:predictReWe_direto_aprox}
\end{figure}

To assess the feasibility of predicting dimensionless numbers using only geometric data, we trained an LSTM network using the collision training dataset. Fig. \ref{fig:predictReWe_direto_aprox} illustrates the predicted values for each instance of the collision test dataset while a detailed metric analysis is presented in Table \ref{tab:metricas_direto}.

According to the results in Figs. \ref{fig:predictReWe_aprox} and \ref{fig:predictReWe_direto_aprox} and analyzing the difference in metrics from Table \ref{tab:metricas_direto}, it is evident that the performance of the one-phase neural network, operating solely on geometrical data as inputs, is inferior to that of our two-stage prediction strategy presented in Section \ref{sec:dn}. 

\renewcommand{\arraystretch}{1.5}
\begin{table}[H]
    \begin{center}
        \begin{tabular}{  |p{2cm}||p{3cm}|p{3cm}|  }
            \hline
                Metric & Two-phase & Direct \\
            \hline
                $R^2$ & 0.99823 & 0.95478 \\
                $RMSE$ & 1.39455 & 7.04277 \\
                $NRMSE$ & 0.0141 & 0.07122 \\
            \hline
        \end{tabular}
        \caption{Comparison of the evaluation metrics for the two-phase and direct approaches.}
        \label{tab:metricas_direto}
    \end{center}
\end{table}
\renewcommand{\arraystretch}{1.0}

\begin{figure}[ht]
    \centering
    \includegraphics[width=0.7\textwidth]{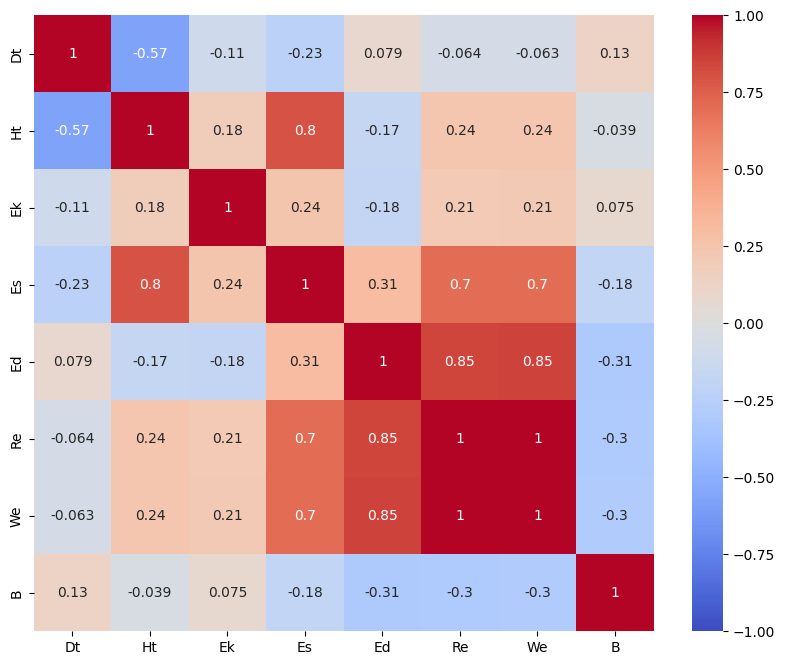}
    \caption{Pearson correlation matrix illustrating the pairwise correlation coefficients between each variable.}
    \label{fig:correlacao}
\end{figure}

This is likely related to stronger correlation between the dimensionless numbers and energy values, as the dimensionless numbers are used in the non-dimensionalization of energy terms. This observation is further supported by the Pearson correlation matrix shown in Fig. \ref{fig:correlacao}, which reveals a high correlation among these terms and a low correlation with the geometric data. This comparison highlights the crucial role that is undertaken by the two-phase sequential neural network approach in enhancing the accuracy and robustness of dimensionless number predictions.

\end{appendices}

\printbibliography

\end{document}